\documentclass[journal, twoside]{IEEEtran}

\usepackage{amsmath, amsfonts}   % From the American Mathematical Society
\usepackage{algorithm,algorithmic}
\usepackage{graphicx}

\newtheorem{theorem}{Theorem}
\newtheorem{definition}{Definition}
\newtheorem{lemma}{Lemma}
\newtheorem{condition}{Condition}
\newtheorem{corollary}{Corollary}
\def\Algorithm#1{{\bf Algorithm~#1}}
\def\mod{\mbox{~mod~}}
\newenvironment{thmproof}[1]
{\noindent\hspace{2em}{\it #1 }} {\hspace*{\fill}~\QED\par\endtrivlist\unskip}
\def \markalgline#1{\newcounter{ctrline#1} \setcounter{ctrline#1}{\value{ALC@line}}}
\def \readalgline#1{\arabic{ctrline#1}}
\def \cd#1{\mbox{\sf Cond#1}}
\def \mdt#1{#1\mbox{~mod~}2}
\def \ctr#1{\mbox{ctr#1}}
\def \st{\mbox{ s.t.~}}
\def \nbd{\mbox{-nbd}}
\def\EE{{\mathsf E}}
\def\cmpt{\hspace{-.02cm}}

\def \Theorem#1{{\it Theorem~#1}}
\def \Theorems#1{{\it Theorems~#1}}

\def \Condition#1{{\it Condition~#1}}

\def \Lemma#1{{\it Lemma~#1}}
\def \Lemmas#1{{\it Lemmas~#1}}

\def \Figure#1{{\rm Fig.~#1}}

\def \Table#1{{\rm TABLE~#1}}

\def \Append#1{{\sc Appendix~#1}}

\begin{document}
%
% paper title
\title{Bandit Problems with Side Observations}
%
%
% author names and IEEE memberships
% note positions of commas and nonbreaking spaces ( ~ ) LaTeX will not break
% a structure at a ~ so this keeps an author's name from being broken across
% two lines.
% use \thanks{} to gain access to the first footnote area
% a separate \thanks must be used for each paragraph as LaTeX2e's \thanks
% was not built to handle multiple paragraphs
\author{Chih-Chun~Wang,~\IEEEmembership{Student Member,~IEEE,}
        Sanjeev~R.~Kulkarni,~\IEEEmembership{Fellow,~IEEE,}
        and~H.~Vincent~Poor,~\IEEEmembership{Fellow,~IEEE}% <-this % stops a space
\thanks{Manuscript received November 15, 2002; revised November 9, 2004.
        This work was supported in part
        by the National Science Foundation under Grants ANI-0338807 and ECS-9873451,
        the Army Research Office under contract number DAAD19-00-1-0466, the Office of
        Naval Research under Grant No.~N00014-03-1-0102, and by the New Jersey Center
        for Pervasive Information Technologies.}% <-this % stops a space
\thanks{C-C.~Wang, S.R.~Kulkarni, and H.V.~Poor are with Princeton University.}}
% note the % following the last \IEEEmembership and also the first \thanks -
% these prevent an unwanted space from occurring between the last author name
% and the end of the author line. i.e., if you had this:
%
% \author{....lastname \thanks{...} \thanks{...} }
%                     ^------------^------------^----Do not want these spaces!
%
% a space would be appended to the last name and could cause every name on that
% line to be shifted left slightly. This is one of those "LaTeX things". For
% instance, "A\textbf{} \textbf{}B" will typeset as "A B" not "AB". If you want
% "AB" then you have to do: "A\textbf{}\textbf{}B"
% \thanks is no different in this regard, so shield the last } of each \thanks
% that ends a line with a % and do not let a space in before the next \thanks.
% Spaces after \IEEEmembership other than the last one are OK (and needed) as
% you are supposed to have spaces between the names. For what it is worth,
% this is a minor point as most people would not even notice if the said evil
% space somehow managed to creep in.
%
% The paper headers
\markboth{IEEE Transactions on Automatic Control,~Vol.~50,
No.~5,~May~2005}{Wang \MakeLowercase{\textit{et al.}}: Bandit
Problems with Side Observations}
% The only time the second header will appear is for the odd numbered pages
% after the title page when using the twoside option.
%
% *** Note that you probably will NOT want to include the author's name in ***
% *** the headers of peer review papers.                                   ***

% If you want to put a publisher's ID mark on the page
% (can leave text blank if you just want to see how the
% text height on the first page will be reduced by IEEE)
%\pubid{0000--0000/00\$00.00~\copyright~2002 IEEE}

% use only for invited papers
%\specialpapernotice{(Invited Paper)}

% make the title area
\maketitle

\begin{abstract}
An extension of the traditional two-armed bandit problem is considered, in
which the decision maker has access to some side information before deciding
which arm to pull.  At each time $t$, before making a selection, the decision
maker is able to observe a random variable $X_t$ that provides some information
on the rewards to be obtained. The focus is on finding uniformly good rules
(that minimize the growth rate of the inferior sampling time) and on
quantifying how much the additional information helps.  Various settings are
considered and for each setting, lower bounds on the achievable inferior
sampling time are developed and asymptotically optimal adaptive schemes
achieving these lower bounds are constructed.
\end{abstract}

% Note that keywords are not normally used for peerreview papers.

% For peer review papers, you can put extra information on the cover
% page as needed:
% \begin{center} \bfseries EDICS Category: 3-BBND \end{center}
%
% For peerreview papers, inserts a page break and creates the second title.
% Will be ignored for other modes.
\IEEEpeerreviewmaketitle

\section{Introduction}
% The very first letter is a 2 line initial drop letter followed
% by the rest of the first word in caps.
%
% form to use if the first word consists of a single letter:
% \PARstart{A}{demo} file is ....
%
% form to use if you need the single drop letter followed by
% normal text (unknown if ever used by IEEE):
% \PARstart{A}{}demo file is ....
%
% Some journals put the first two words in caps:
% \PARstart{T}{his demo} file is ....
%
% Here we have the typical use of a "T" for an initial drop letter
% and "HIS" in caps to complete the first word.
\PARstart{S}{ince} the publication of \cite{Robbins52}, bandit problems have
attracted much attention in various areas of statistics, control, learning, and
economics (e.g., see
\cite{Adam01,Berry72,Chernoff72,GhoshSen91,Gittins79a,Gittins79b,LaiRobbins84,LaiRobbins85,LaiYakowitz95}).
In the classical two-armed bandit problem, at each time a player selects one of
two arms and receives a reward drawn from a distribution associated with the
arm selected. The essence of the bandit problem is that the reward
distributions are unknown, and so there is a fundamental trade-off between
gathering information about the unknown reward distributions and choosing the
arm we currently think is the best. A rich set of problems arises in trying to
find an optimal/reasonable balance between these conflicting objectives (also
referred to as learning versus control, or exploration versus exploitation).

We let $\{Y^1_\tau\}$ and $\{Y^2_\tau\}$ denote the sequences of rewards from
arms~1 and 2 in a two-armed bandit machine. In the traditional parametric
setting, the underlying configurations/distributions of the arms are expressed
by a pair of parameters $C_0=(\theta_1, \theta_2)$ such that $\{Y^1_\tau\}$ and
$\{Y^2_\tau\}$ are independent and identically distributed (i.i.d.) with
distribution $(F_{\theta_1},F_{\theta_2})$, where $\{F_\theta\}$ is a known
family of distributions parametrized by $\theta$. The goal is to maximize the
sum of the expected rewards. Results on achievable performance have been
obtained for a number of variations and extensions of the basic problem defined
in \cite{LaiRobbins85} (e.g., see
\cite{AgrawalHegdeTeneketzis88,AgrawalTeneketzisAnantharam89a,AgrawalTeneketzisAnantharam89b,AnantharamVaraiyaWalrand87a,AnantharamVaraiyaWalrand87b,KatehakisRobbins95,KulkarniLugosi00}).

In this paper, we consider an extension of the classical two-armed bandit where
we have access to  side information before making our decision
 about which arm to pull.  Suppose at time~$t$, in
addition to the history of previous decisions, outcomes, and observations, we
have access to a side observation $X_t$ to help us make our current decision.
The extent to which this side observation can help depends on the relationship
of $X_t$ to the reward distributions of $Y^1_t$ and $Y^2_t$.
%, namely the
%configuration $(\theta_1,\theta_2)$ and the sequence of rewards $Y^i_t$.

Previous work on bandit problems with side observations includes
\cite{Clayton89,Kulkarni93,Sarkar91,Woodroofe79,Zoubeidi94}. Woodroofe
\cite{Woodroofe79} considered a one-armed bandit in a Bayesian setting, and
constructed a simple criterion for asymptotically optimal rules. Sarkar
\cite{Sarkar91} extended the side information model of \cite{Woodroofe79}  to
the exponential family. In \cite{Kulkarni93}, Kulkarni considered  classes of
reward distributions and their effects on performance using results from
learning theory. Most of the previous work with side observations is on
one-armed bandit problems, which can be viewed as a special case of the
two-armed setting by letting arm~2 always return zero.

%In \cite{Clayton89},  with the help of the ``link'' function, an index-based
%rule is provided. In \cite{Zoubeidi94}, side information was used as the index
%of different categories.

In contrast with this previous work, we consider various general settings of
side information for a two-armed bandit problem.  Our focus is on providing
both lower bounds and bound-achieving algorithms for the various settings. The
results and proofs are very much along the lines of \cite{LaiRobbins84} and
subsequent works as in
\cite{AgrawalHegdeTeneketzis88,AgrawalTeneketzisAnantharam89a,AgrawalTeneketzisAnantharam89b,AnantharamVaraiyaWalrand87a,AnantharamVaraiyaWalrand87b}.

We now describe the settings considered in this paper.
\begin{enumerate}
\item {\bf Direct Information:} In this case, $X_t$ provides
information directly about the underlying configuration $C_0=(\theta_1,
\theta_2)$, which  allows a type of separation between the learning and
control. This has a dramatic effect on the achievable inferior sampling time.
Specifically, estimating $(\theta_1,\theta_2)$ by observing $\{X_\tau\}$, and
using the estimate $(\hat{\theta}_1,\hat{\theta}_2)$ to make the decision,
results in bounded expected inferior sampling time.
\end{enumerate}

If the distribution of $\{X_\tau\}$ is not a function  of $C_0$, we are not
able to learn $C_0$ through $\{X_\tau\}$. However, different values of the side
observation $X_t$ will result in different conditional distributions of the
rewards $Y^i_t$. By exploiting this new structure (observing $X_t$ in advance),
we can hope to do better than the case without any side observation.

A physical meaning about the above scenario (constant distribution on
$\{X_\tau\}$) is that a two-armed bandit with the side observations drawn from
a finite set $\{x_1,x_2,\cdots,x_n\}$ can be viewed as a set of $n$ different
two-armed sub-bandit machines indexed from $x_1$ to $x_n$. The player does not
know the order of sub-machines he is going to play, which is determined by
rolling a die with $n$ faces. However, by observing $X_t$, the player knows
which machine (out of the $n$ different ones) he is facing now before selecting
which arm to play. The connection between these sub-machines is that they share
the same common configuration pair $(\theta_1,\theta_2)$, so that the rewards
observed from one machine provide information on the common
$(\theta_1,\theta_2)$, which can then be applied to {\em all} of the others
(different values of $X_t$). This is the key aspect that makes this setup
distinct from simply having many independent bandit problems with random access
opportunity.

%It is worth noting that within this setting, the most rewarding arm at time $t$
%is in general a function of both the underlying configuration pair $C_0$ and
%the side observation $X_t$.

We consider the following three cases of different relationships among the most
rewarding arm, $C_0$, and $X_t$.

\begin{enumerate}
\setcounter{enumi}{1}
\item {\bf For all possible $C_0$, the best arm is a function of $X_t$:}
That is, $\forall (\theta_1,\theta_2), \exists x_1,x_2$ such that at time~$t$,
arm~1 yields higher expected reward conditioned on $X_t=x_1$ while arm~2 is
preferred when $X_t=x_2$. Surprisingly, we exhibit an algorithm that achieves
{\em bounded} expected inferior sampling time in this case. Woodroofe's result
\cite{Woodroofe79} can then be viewed as a special case of this scenario.

\item {\bf For all possible $C_0$, the best arm is not a function of $X_t$:}
In this case, for {\em all} configurations $(\theta_1,\theta_2)$, one of the
arms is always preferred regardless of the value of $X_t$.  Since the
conditional reward distributions are functions of  $X_t$, the intuition is that
we can postpone our learning until it is most advantageous to us. We show that,
asymptotically, our performance will be governed by the most ``informative''
bandit (among the different values taken on by $X_t$).
%Hence, we pay no penalty in
%terms of the constant in the $\log t$ growth rate of the inferior sampling time
%for having to learn which is the most advantageous bandit.

\item {\bf Mixed Case:}
This is a general case that combines the previous two, and contains the main
contribution of this paper.  For some possible configurations, one arm may
always be preferred (for any $X_t$), while for other possible configurations,
the preferred arm is a function of $X_t$.  We exhibit an algorithm that
achieves the best possible in either case.  That is, if the best arm is a
function of  $X_t$, it achieves bounded expected inferior sampling time as in
Case 2, while if the underlying configuration is such that one arm is always
preferred, then we get the results of Case 3.

\end{enumerate}

Our paper is organized as follows. In Section \ref{sec: formulation}, we
introduce the general formulation.  In Section \ref{sec:traditional-bandits},
we provide background on the asymptotic analysis of traditional bandit problems
(without side observations). In Sections \ref{sec: d-info} through \ref{sec:
mixed-case}, we consider the above four cases respectively. The results are
included in each section, while details of the proofs are provided in the
appendix.
\section{General Formulation\label{sec: formulation}}
Consider the two-armed bandit problem defined as follows.  Suppose we have two
sequences of (real-valued) random variables (r.v.'s), $\{Y^i_\tau\}_{i=1,2}$,
and an i.i.d.\ side observation sequence $\{X_\tau\}$, taking values in
${\mathbf X}\subset{\mathbb R}$. $\{Y^i_\tau\}$ denotes the reward sequence of
arm $i$ while $X_t$ is the side information observed at time $t$ before making
the decision. The formal parametric setting is as follows. For each
configuration pair $C_0=(\theta_1,\theta_2)$ and each~$i$, the sequence of
vectors $(X_t, Y^i_t)$ is i.i.d.\ with joint distribution
$G_{C_0}(dx)F_{\theta_i}(dy|x)$, where the families $\{G_C\}_{C\in{\mathbf
\Theta}^2}$ and $\{F_\theta(\cdot|\cdot)\}_{\theta\in{\mathbf \Theta}}$ are
known to the player, but the true value of the corresponding index $C_0$ must
be learned through experiments. For notational simplicity, we further assumed
${\mathbf{\Theta}}$ is a set of real numbers.

\def\tnotn#1{\parbox[t]{1.2cm}{\centering#1}}
\def\tdesc#1{\parbox[t]{6.5cm}{#1}}
\begin{table}[t]
    \caption{Glossary}
    \label{tab: glossary}
\begin{center}{
    \begin{tabular}{cl}
        \hline
        Not'n & Description\\
        \hline
%        \tnotn{${\mathbf\Theta}, {\mathbf \Theta}^2$} & \tdesc{${\mathbf{\Theta}}\subseteq{\mathbb{R}}$ is the set of all possible $\theta$; ${\mathbf \Theta}^2$ is the set of parameter pairs.}\\
        \tnotn{$G_{C}(dx)$}& \tdesc{The marginal distribution of the i.i.d.\ $\{X_\tau\}$ under configuration $C$.}\\
        \tnotn{$F_{\theta_i}(dy|x)$} & \tdesc{The conditional distribution of the reward of arm $i$, $Y^i_t$, under parameter $\theta_i$.}\\
        \tnotn{$\mu_{\theta}(x)$} & \tdesc{The conditional expectation of the reward, $\mu_{\theta}(x)={\mathsf E}_\theta\{Y|x\}=\int yF_\theta(dy|x)$.}\\
        \tnotn{$1(C_0)$, $2(C_0)$} & \tdesc{The first and the second coordinates of the configuration pair $C_0$,
            i.e.\ $1(C_0)=\theta_1$, $2(C_0)=\theta_2$. For example: $F_{1(C_0)}(dy|x)=F_{\theta_1}(dy|x)$ and $\mu_{2(C_0)}(x)=\mu_{\theta_2}(x)$.}\\
        \tnotn{$M_C(x)$} & \tdesc{The index of the preferred arm, i.e.\ $\arg\max_{i=1,2} \{\mu_{i(C)}(x)\}$.}\\
        \tnotn{$\phi_t$} & \tdesc{The decision rule taking values in $\{1,2\}$ and depending only on
                    the past outcomes and the current side information $X_t$.}\\
        \tnotn{$T_i(t)$} & \tdesc{The total number of samples taken on arm $i$ up to time~$t$,\\
                $T_i(t)=\sum^t_{\tau=1}1_{\{\phi_\tau=i\}}$.}\\
        \tnotn{$T_{inf}(t)$} & \tdesc{The total number of samples taken on the inferior arm up to
                time~$t$: $T_{inf}(t)=\sum^t_{\tau=1}1_{\{\phi_\tau\neq M_{C_0}(X_\tau)\}}$.}\\
        \tnotn{$I(P, Q)$} & \tdesc{The Kullback-Leibler (K-L) information number between distributions $P$ and
        $Q$:
              $I(P, Q)={\mathsf E}_P\left\{\log\left(        \frac{dP}{dQ} \right)\right\}$.}\\
        \tnotn{$I(\theta_1, \theta_2|x)$} & \tdesc{The conditional K-L information number: $I(\theta_1, \theta_2|x)=I(F_{\theta_1}(\cdot|x),
        F_{\theta_2}(\cdot|x))$.}\\
        \hline
    \end{tabular}
    }
\end{center}
\end{table}

Note that the concept of the i.i.d.\ bandit is now extended to the assumption
that the vector sequence $\{(X_t, Y^i_t)\}_t$ is  i.i.d. The unconditioned
marginal sequence $\{Y^i_\tau\}$ remains i.i.d. However, rather than the
unconditional marginals,  the player is now facing the conditional distribution
of $Y^i_t$, which is a function of the observed side information $X_t$ (and is
not identically distributed given different $X_t$).

 The goal is to find
an adaptive allocation rule $\{\phi_\tau\}$ to maximize the growth rate of the
expected reward:
\begin{eqnarray}
{\mathsf E}_{C_0}\{W_\phi(t)\}:={\mathsf E}_{C_0}\left\{ \sum^t_{\tau=1}\left(
                1_{\{\phi_\tau=1\}}Y^1_\tau+
                1_{\{\phi_\tau=2\}}Y^2_\tau\right)
\right\},\nonumber
\end{eqnarray}
or equivalently to minimize the growth rate of the expected inferior sampling
time\footnotemark, namely ${\mathsf E}_{C_0}\{T_{inf}(t)\}$. To be more
explicit, at any time $t$, $\phi_t$ takes a value in $\{1,2\}$ and depends only
on the past rewards ($\tau<t$)
 and the current side observation $X_t$.

\footnotetext{In the literature of bandit problems, the term ``regret" is more
typically used rather than the inferior sampling time. For traditional
two-armed bandits, the regret is defined as
\begin{eqnarray}
\mbox{regret}:=t\cdot\max\{\mu_{\theta_1}, \mu_{\theta_2}\}-{\mathsf
E}_{C_0}\{W_\phi(t)\},\nonumber
\end{eqnarray}
the difference between the best possible reward and that of the strategy of
interest $\{\phi_\tau\}$. The relationship between the regret and $T_{inf}(t)$
is as follows.
\begin{eqnarray}
\mbox{regret}=|\mu_{\theta_1}-\mu_{\theta_2}|\cdot{\mathsf
E}_{C_0}\{T_{inf}(t)\}.\nonumber
\end{eqnarray}
For greater simplicity in the discussion of bandit problems with side
observations, we consider $T_{inf}(t)$ rather than the regret.}

We define a uniformly good rule as follows.
\begin{definition}[Uniformly Good Rules]
An allocation rule $\{\phi_\tau\}$ is uniformly good if for all $C\in{\mathbf
\Theta}^2$, ${\mathsf E}_{C}\{T_{inf}(t)\}=o(t^\alpha)$, $\forall \alpha>0$.
\end{definition}
In what follows, we consider only uniformly good rules and regard other rules
as uninteresting. Necessary notation and several quantities of interest are
defined in \Table{\ref{tab: glossary}}. We assume that all the given
expectations exist and are finite.

%%%%%%%%%%%%%%%%%%%%%%%%%%%%%%%%%%%%%%%%%%%%%%%%%%%%%%%%%%%%%%%%%%%%%%%%%%%%%%%%%%%%%%%%%%
%
%   Traditional Bandit
%
%%%%%%%%%%%%%%%%%%%%%%%%%%%%%%%%%%%%%%%%%%%%%%%%%%%%%%%%%%%%%%%%%%%%%%%%%%%%%%%%%%%%%%%%%%
\section{Traditional Bandits\label{sec:traditional-bandits}}
Under the general formulation provided in Section \ref{sec: formulation}, the
traditional non-Bayesian, parametric, infinite horizon, two-armed bandit is
simply a degenerate case, i.e., the traditional bandit problem is equivalent to
having only one element in $\mathbf X$ (say ${\mathbf X}=\{x_0\}$). This
formulation of traditional bandit problems is identical to the two-armed case
of \cite{AnantharamVaraiyaWalrand87a,LaiRobbins84,LaiRobbins85}. For
simplicity, the argument $x_0$ can be omitted in this traditional setting,
i.e., $M_C:=M_C(x_0)$, $\mu_\theta:=\mu_\theta(x_0)$,
$I(\theta_1,\theta_2):=I(\theta_1,\theta_2|x_0)$, etc.

The main contribution of
\cite{AnantharamVaraiyaWalrand87a,LaiRobbins84,LaiRobbins85} is the asymptotic
analysis stated as the following two theorems.
\begin{theorem}[$\log t$ Lower Bound]
For any uniformly good rule $\{\phi_\tau\}$, $T_{inf}(t)$ satisfies
\begin{eqnarray}
        &&\lim_{t\rightarrow\infty}{\mathsf P}_{C_0}\left(
                T_{inf}(t)\geq\frac{(1-\epsilon)\log t}{K_{C_0}}
        \right)=1,~\forall\epsilon>0,\nonumber\\
\mbox{and~~}        &&\liminf_{t\rightarrow\infty}\frac{{\mathsf
E}_{C_0}\{T_{inf}(t)\}}{\log t}\geq
        \frac{1}{K_{C_0}},\nonumber%\label{eq:basic-theorem}
\end{eqnarray}
where $K_{C_0}$ is a constant depending on $C_0$. If $M_{C_0}=2$, then
$T_{inf}(t)=T_1(t)$ and $K_{C_0}$ is defined\footnote{Throughout this paper, we
will adopt the conventions that the infimum of the null set is $\infty$, and
$\frac{1}{\infty}=0$.} as follows.
\begin{eqnarray}
K_{C_0}=\inf\{I(\theta_1, \theta):\forall \theta,
\mu_\theta>\mu_{\theta_2}\}.\label{eq:def-KC0}
\end{eqnarray}
The expression for $K_{C_0}$ for the case in which $M_{C_0}=1$ can be obtained
by symmetry. \label{thm:basic-theorem}
\end{theorem}
\begin{theorem}[Asymptotic Tightness]
Under certain regularity conditions\footnote{If the parameter set is finite,
\Theorem{\ref{thm:basic-theorem2}} always holds. If ${\mathbf \Theta}$ is the
set of reals, the required regularity conditions are on the unboundedness and
the continuity of $\mu_\theta$ w.r.t.\ $\theta$ and on the continuity of
$I(\theta_1,\theta)$ w.r.t.\ $\mu_\theta$. }, the above lower bound is
asymptotically tight. Formally stated, given the distribution family
$\{F_{\theta}\}$, there exists a decision rule $\{\phi_\tau\}$ such that for
all $C_0=(\theta_1,\theta_2)\in{\mathbf \Theta}^2$,
\begin{eqnarray}
        \limsup_{t\rightarrow\infty}\frac{{\mathsf E}_{C_0}\{T_{inf}(t)\}}{\log t}\leq
        \frac{1}{K_{C_0}},\nonumber
\end{eqnarray}
where $K_{C_0}$ is the same as in
\Theorem{\ref{thm:basic-theorem}}.\label{thm:basic-theorem2}
\end{theorem}

The intuition behind the $\log t$ lower bound is as follows. Suppose
$M_{C_0}=2$ and consider another configuration $C'=(\theta,\theta_2)$ such that
$M_{C'}=1$. It can be shown that if under configuration
$C_0=(\theta_1,\theta_2)$, ${\mathsf E}_{C_0}\{T_{inf}(t)\}$ is less than the
$\log t$ lower bound,  ${\mathsf E}_{C'}\{T_{inf}(t)\}$ must be greater than
$o(t^\alpha)$ for some $\alpha>0$, which contradicts the assumption that
$\{\phi_\tau\}$ is uniformly good.
%The detailed proof,  presented in
%Appendix~\ref{app: lbd-X-depends-on-x} for a more general setting,  is similar
%to the proof of Stein's lemma in hypothesis testing between $H_0$ and $H_1$,
%the change-of-measure argument between ${\mathsf P}_1(\mbox{accept $H_0$})$ and
%${\mathsf P}_0(\mbox{accept $H_0$})$.

%Note that under configuration $C_0$, $T_{inf}(t)=T_1(t)$, while
%under configuration $C'$, $T_{inf}(t)=T_2(t)$.

%%%%%%%%%%%%%%%%%%%%%%%%%%%%%%%%%%%%%%%%%%%%%%%%%%%%%%%%%%%%%%%%%%%%%%%%%%%%%%%%%%%%%%%%%%
%
%   Direct information
%
%%%%%%%%%%%%%%%%%%%%%%%%%%%%%%%%%%%%%%%%%%%%%%%%%%%%%%%%%%%%%%%%%%%%%%%%%%%%%%%%%%%%%%%%%%
\section{Direct Information\label{sec: d-info}}
%%%%%%%%%%%%%%%%%%%%%%%%%%%%%%%%%%%%%%%%%%%%%%%%%%%%%%%%%%%%%%%%%%%%%%%%%%%%%%%%%%%%%%%%%%
%
%   Direct information: formulation
%
%%%%%%%%%%%%%%%%%%%%%%%%%%%%%%%%%%%%%%%%%%%%%%%%%%%%%%%%%%%%%%%%%%%%%%%%%%%%%%%%%%%%%%%%%%
\subsection{Formulation}
In this setting, the side observation $X_t$ directly reveals information about
the underlying configuration pair $C_0=(\theta_1,\theta_2)$ in the following
way.
\begin{description}
\item [Dependence:]~~~~~~~~ $G_{C_1}=G_{C_2}$ iff $C_1=C_2$.
\end{description}
As a result, observing the empirical distribution of $X_t$ gives us useful
information about the underlying parameter pair $C_0$. Thus this is a type of
identifiability condition.

\indent Examples:
\begin{itemize}
\item ${\mathbf \Theta}=(0,0.5)$ and ${\mathbf X}=\{x_1,x_2,x_3\}$.
\begin{eqnarray}
{\mathsf P}_{(\theta_1,\theta_2)}(X_t=x_k)=\begin{cases} \theta_k& \text{if $k=1,2$} \\
1-\theta_1-\theta_2& \text{otherwise}
\end{cases}.\nonumber
\end{eqnarray}
\item ${\mathbf \Theta}=(0,\infty)$ and ${\mathbf X}=[0,1]$. $X_t$ is beta distributed with parameters $(\theta_1,\theta_2)$.
\end{itemize}

%%%%%%%%%%%%%%%%%%%%%%%%%%%%%%%%%%%%%%%%%%%%%%%%%%%%%%%%%%%%%%%%%%%%%%%%%%%%%%%%%%%%%%%%%%
%
%   Direct information: bounded scheme
%
%%%%%%%%%%%%%%%%%%%%%%%%%%%%%%%%%%%%%%%%%%%%%%%%%%%%%%%%%%%%%%%%%%%%%%%%%%%%%%%%%%%%%%%%%%
\subsection{Scheme with Bounded ${\mathsf E}_{C_0}\{T_{inf}(t)\}$\label{subsec: d-info-bdd-regret}}
Consider the following condition.
\begin{condition} \label{cond: d-info}For any fixed $C_0$
% and any estimate sequence $\{\hat{C}_\tau\}$, such that
% $\hat{C}_t\rightarrow C_0$, there exists
%$t_0$ such that for all $x\in{\mathbf X}$ and $t>t_0$,
%$M_{\hat{C}_t}(x)=M_{C_0}(x)$. Or equivalently,
\begin{eqnarray}
\inf\left\{\rho(G_{C_0}, G_{C_e}):C_e\in{\mathbf \Theta^2}, \exists x,
M_{C_e}(x)\neq M_{C_0}(x)\right\}>0,\nonumber
\end{eqnarray}
where $\rho$ denotes the Prohorov metric\footnote{A definition of the Prohorov
metric is stated in \Append{\ref{subsec:sanov-prohorov}}.} on the space of
distributions. Two examples satisfying \Condition{\ref{cond: d-info}} are as
follows:
\end{condition}
%Here we provide two examples satisfying {\it Condition~\ref{cond: d-info}}:
\begin{itemize}
\item {\em Example 1:} $\mathbf X$ is finite, and $\forall x\in\mathbf
X$, $\mu_\theta(x)$ is continuous with respect to (w.r.t.)~$\theta$.
\item {\em Example 2:} $F_\theta(\cdot|x)\sim{\mathcal N}(\theta
x,1)$ is a Gaussian distribution with mean $\theta x$ and variance $1$.
\end{itemize}

Under this condition, we obtain the following result.
\begin{theorem}[Bounded ${\mathsf E}_{C_0}\{T_{inf}(t)\}$]\label{thm: d-info}
If \Condition{\ref{cond: d-info}} is satisfied, then there exists an allocation
rule $\{\phi_\tau\}$, such that $\lim_{t\rightarrow\infty}{\mathsf
E}_{C_0}\{T_{inf}(t)\}<\infty$ and $\lim_{t\rightarrow\infty}T_{inf}(t)<\infty$
a.s.
\end{theorem}
\begin{itemize}
\item Note: the information directly revealed by $X_t$ helps the
sequential control scheme  surpass the $\log t$ lower bound stated in
\Theorem{\ref{thm:basic-theorem}}. This significant improvement (bounded
expected inferior sampling time) is due to the fact that the dilemma between
learning and control no longer exists in the direct information case.
\end{itemize}

%%%%%%%%%%%%%%%%%%%%%%%%%%%%%%%%%%%%%%%%%%%%%%%%%%%%%%%%%%%%%%%%%%%%%%%%%%%%%%%%%%%%%%%%%%
%
%   Direct information: bounded scheme, Intuition
%
%%%%%%%%%%%%%%%%%%%%%%%%%%%%%%%%%%%%%%%%%%%%%%%%%%%%%%%%%%%%%%%%%%%%%%%%%%%%%%%%%%%%%%%%%%
%Condition \ref{cond: d-info} implies that whenever our estimate, $\hat{C}_t$, is close enough to
%$C_0=(\theta_1,\theta_2)$, then $\forall x\in{\mathbf X}$, the preference order of
%$\mu_{1(\hat{C}_t)}(x), \mu_{2(\hat{C}_t)}(x)$, is the same as $\mu_1(x),\mu_2(x)$. So the question
%is reduced to how fast we can have good enough estimates.

%%%%%%%%%%%%%%%%%%%%%%%%%%%%%%%%%%%%%%%%%%%%%%%%%%%%%%%%%%%%%%%%%%%%%%%%%%%%%%%%%%%%%%%%%%
%
%   Direct information: The SCHEME
%
%%%%%%%%%%%%%%%%%%%%%%%%%%%%%%%%%%%%%%%%%%%%%%%%%%%%%%%%%%%%%%%%%%%%%%%%%%%%%%%%%%%%%%%%%%

We provide a scheme achieving bounded ${\mathsf E}_{C_0}\{T_{inf}(t)\}$ as in
\Algorithm{\ref{alg:direct-info}}, of which a detailed analysis is given in
\Append{\ref{app: pf-d-info}}.
\begin{algorithm}[h]\footnotesize
\caption{$\phi_{t}$, the decision at time $t$ (after observing $X_t$ but before
deciding $\phi_t$) }\label{alg:direct-info}
\begin{algorithmic}[1]
\STATE Construct
    \begin{eqnarray}
        {\mathbf C}_t:=
        \left\{
            C\in {\mathbf \Theta}^2: \rho(G_C, L_X(t))\leq\inf_{C\in{\mathbf \Theta}^2}\rho(G_C,
            L_X(t))+\frac{1}{t}
        \right\},\nonumber
    \end{eqnarray}
    where $L_X(t)$ is the empirical measure of the side observations $\{X_\tau\}$ until time $t$, and $\rho$~is the
    Prohorov metric as before.

\STATE Arbitrarily pick $\hat{C}_t\in {\mathbf C}_t$, and set
$\phi_{t}=M_{\hat{C}_t}(X_{t})$.
\end{algorithmic}
\end{algorithm}

%%%%%%%%%%%%%%%%%%%%%%%%%%%%%%%%%%%%%%%%%%%%%%%%%%%%%%%%%%%%%%%%%%%%%%%%%%%%%%%%%%%%%%%%%%
%
%   Expecting the best
%
%%%%%%%%%%%%%%%%%%%%%%%%%%%%%%%%%%%%%%%%%%%%%%%%%%%%%%%%%%%%%%%%%%%%%%%%%%%%%%%%%%%%%%%%%%
\section{Best Arm As A Function Of $X_t$\label{sec: depends-on-x}}
%%%%%%%%%%%%%%%%%%%%%%%%%%%%%%%%%%%%%%%%%%%%%%%%%%%%%%%%%%%%%%%%%%%%%%%%%%%%%%%%%%%%%%%%%%
%
%   Expecting the best: formulations
%
%%%%%%%%%%%%%%%%%%%%%%%%%%%%%%%%%%%%%%%%%%%%%%%%%%%%%%%%%%%%%%%%%%%%%%%%%%%%%%%%%%%%%%%%%%
For all of the following sections (Sections~\ref{sec: depends-on-x} through
\ref{sec: mixed-case}), we consider only the case in which observing $X_t$ will
not reveal any information about $C_0$, but only reveals information about the
upcoming reward $Y^i_t$, that is,
\begin{itemize}
\item $G_{C_0}$ does not depend on the value of $C_0$;  we use
$G:=G_{C_0}$ as shorthand notation.
\end{itemize}

Three further refinements regarding the relationship between $M_C(x)$ and $x$
will be discussed separately (each in one section).
\subsection{Formulation}
\begin{figure}[t]\centering
\includegraphics[width=2in, keepaspectratio=true]{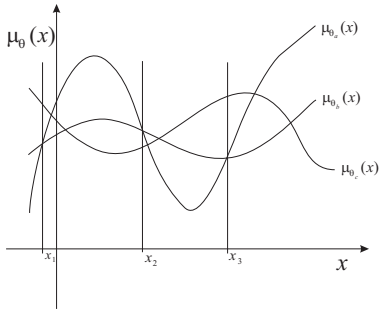}
\caption{The best arm at time $t$ {\em always} depends on the side observation
$X_t$. That is, for any possible pair $(\theta_1,\theta_2)$ the two curves,
$\mu_{\theta_1}(x)$ and $\mu_{\theta_2}(x)$, (w.r.t.\ $x$) always intersect
each other. } \label{fig1}
\end{figure}

In this section, we assume that for all possible $C$, the side observation
$X_t$ is {\em always} able to change the preference order as shown in
\Figure{\ref{fig1}}. That is,
\begin{itemize}
\item For all $C\in{\mathbf \Theta}^2$,  there exist $x_1$ and $x_2$ such that
$M_C(x_1)=1$ and $M_C(x_2)=2$.
\end{itemize}

The needed regularity conditions are as follows.
\begin{enumerate}
\item ${\mathbf X}$ is a finite set and ${\mathsf P}_G(X_t=x)>0$ for all $x\in{\mathbf X}$.
\item $\forall \theta_1,\theta_2,x$, $I(\theta_1,\theta_2|x)$ is strictly positive  and finite.
\item $\forall x$, $\mu_\theta(x)$ is continuous w.r.t.\ $\theta$.
\end{enumerate}
The first condition embodies the idea of treating $X_t$ as the index of several
different bandit machines, which also simplifies our proof. The second
condition is to ensure that all these different bandit problems are
non-trivial, with {\em non-identical}  pairs of arms.

\indent Example:
\begin{itemize}
\item ${\mathbf \Theta}=(0,\infty)$, ${\mathbf X}=\{-1,1\}$, and the conditional reward  distribution $F_\theta(\cdot|x)\sim {\mathcal N}(\theta
x,1)$.
\end{itemize}

%%%%%%%%%%%%%%%%%%%%%%%%%%%%%%%%%%%%%%%%%%%%%%%%%%%%%%%%%%%%%%%%%%%%%%%%%%%%%%%%%%%%%%%%%%
%
%   Expecting the best: scheme of bounded regret
%
%%%%%%%%%%%%%%%%%%%%%%%%%%%%%%%%%%%%%%%%%%%%%%%%%%%%%%%%%%%%%%%%%%%%%%%%%%%%%%%%%%%%%%%%%%
\subsection{Scheme with Bounded ${\mathsf E}_{C_0}\{T_{inf}(t)\}$}
\begin{theorem}[Bounded ${\mathsf E}_{C_0}\{T_{inf}(t)\}$]
If the above conditions are satisfied, there exists an allocation rule
$\{\phi_\tau\}$ such that
\begin{eqnarray}
\lim_{t\rightarrow\infty}{\mathsf E}_{C_0}\{T_{inf}(t)\} <\infty.\nonumber
\end{eqnarray}
Such a rule is obviously uniformly good. \label{thm: depends-on-x}
\end{theorem}
\begin{itemize}
\item Note: although the side observation $X_t$ does not reveal any
information about $C_0$ in this setting,  the  alternation of the best arm as
the i.i.d.\  $X_t$ takes on different values~$x$ makes it possible to always
perform the control part, $\phi_{t}=M_{\hat{C}_{t-1}}(X_{t})$, and
simultaneously sample both arms often enough. Since the information about both
arms will be implicitly revealed (through the alternation of $M_{C_0}(X_t)$),
the dilemma of learning and control no longer exists, and a significant
improvement ($\lim_{t\rightarrow\infty}{\mathsf E}_{C_0}\{T_{inf}(t)\}<\infty$)
is obtained over the $\log t$ lower bound in \Theorem{\ref{thm:basic-theorem}}.
\end{itemize}

%%%%%%%%%%%%%%%%%%%%%%%%%%%%%%%%%%%%%%%%%%%%%%%%%%%%%%%%%%%%%%%%%%%%%%%%%%%%%%%%%%%%%%%%%%
%
%   Expecting the best: intuition
%
%%%%%%%%%%%%%%%%%%%%%%%%%%%%%%%%%%%%%%%%%%%%%%%%%%%%%%%%%%%%%%%%%%%%%%%%%%%%%%%%%%%%%%%%%%
%\noindent\textit{Intuition:} We introduce an initial forced sampling mechanism that ensures that we
%pull both arms a number of times that has order of growth greater than $O(t^{\frac{1}{2}})$. If $C$
%satisfies the required conditions, we need only to sample the arm we currently think is best
%(depending on $X_t$), and the regular appearances of all $X_t=x^h$ ensure that we sample both arms
%enough (on the order $O(t)$).  As a result, the forced sampling mechanism will terminate quite
%soon, which implies the expected inferior sampling time is bounded.
%%%%%%%%%%%%%%%%%%%%%%%%%%%%%%%%%%%%%%%%%%%%%%%%%%%%%%%%%%%%%%%%%%%%%%%%%%%%%%%%%%%%%%%%%%
%
%   Expecting the best: description of the scheme
%
%%%%%%%%%%%%%%%%%%%%%%%%%%%%%%%%%%%%%%%%%%%%%%%%%%%%%%%%%%%%%%%%%%%%%%%%%%%%%%%%%%%%%%%%%%

\begin{algorithm}[t]\footnotesize
 \caption{$\phi_{t}$, the decision at time $t+1$}\label{alg:depend-on-X} {\bf
Variables:} Denote $T_i^x(t)$ as the total number of time instants until time
$t$ when arm~$i$ has been pulled and  $X_\tau=x$, i.e.\
\begin{eqnarray}
T_i^x(t):=\sum_{\tau=1}^t 1_{\{ X_\tau=x,\phi_\tau=i\}},\nonumber
\end{eqnarray}
and define $x^\star_i:=\arg\max_x \left\{T_i^x(t)\right\}$ and
$T^{x^\star}_i(t):=\max_x\left\{ T_i^x(t)\right\}$.

Construct
\begin{eqnarray}
{\mathbf C}_t&:=&\{ C=(\theta_1,\theta_2)\in{\mathbf \Theta}^2 :\nonumber\\
&&~~\sigma(C,t)\leq\inf\{\sigma(C,t):C\in{\mathbf
\Theta}^2\}+\frac{1}{t}\},\nonumber
\end{eqnarray}
\begin{eqnarray}
\mbox{with~~~~~~~~~~}\sigma(C,t)&:=&\rho(F_{1(C)}(\cdot|x^\star_1),
L_1^{x^\star}(t)),~~~~~~~~~~~~~~~~~~~~~~~~~~\nonumber\\
&&+\rho(F_{2(C)}(\cdot|x_2^\star), L_2^{x^\star}(t)),\nonumber
\end{eqnarray}
where $L_i^x(t)$ is the empirical measure of rewards sampled from arm $i$ at
those time instants $\tau\leq t$ when $X_\tau=x$. (As before $\rho(P,Q)$ is the
Prohorov metric.) Arbitrarily choose $\hat{C}_t\in{\mathbf C}_t$.

%\rule[0pt]{14.5cm}{.5pt}
\hrulefill

\vspace{-.1cm}

{\bf Algorithm:}
\begin{algorithmic}[1]
\IF{$t+1\leq 6$}\markalgline{alg_line:safe_cond}

\STATE $\phi_{t+1}=(t \mod 2)+1$. %(Namely, $\phi_1=1$,
%$\phi_2=2$, $\phi_3=1$, $\phi_4=2$, $\phi_5=1$, $\phi_6=2$.)

\ELSIF{$\exists i$ such that $T_i(t)<\sqrt{t+1}$}

\STATE $\phi_{t+1}=i$.

\ELSE

\STATE   $\phi_{t+1}=M_{\hat{C}_t}(X_{t+1})$.

\ENDIF

\end{algorithmic}
(Note that Line~\readalgline{alg_line:safe_cond} guarantees that there is only
one $i$ such that $T_i(t)<\sqrt{t+1}$.)
\end{algorithm}

%%%%%%%%%%%%%%%%%%%%%%%%%%%%%%%%%%%%%%%%%%%%%%%%%%%%%%%%%%%%%%%%%%%%%%%%%%%%%%%%%%%%%%%%%%
%
%   Expecting the best: The outline of the proof
%
%%%%%%%%%%%%%%%%%%%%%%%%%%%%%%%%%%%%%%%%%%%%%%%%%%%%%%%%%%%%%%%%%%%%%%%%%%%%%%%%%%%%%%%%%%
We construct an allocation rule with bounded ${\mathsf E}_{C_0}\{T_{inf}(t)\}$
given as \Algorithm{\ref{alg:depend-on-X}}. The intuition as to why the
proposed scheme has bounded ${\mathsf E}_{C_0}\{T_{inf}(t)\}$ is as follows.
The forced sampling, $T_i(t)<\sqrt{t+1}$, ensures there are enough samples on
both arms, which implies good enough estimates of $C_0$. Based on the good
enough estimates, the myopic action of sampling the seemingly better arm,
$\phi_{t+1}=M_{\hat{C}_t}(X_{t+1})$, will result in very few inferior
samplings. Unlike the traditional two-armed bandits, in this scenario, the best
arm $M_{C_0}(x)$ varies from one outcome of $X_t$ to the other. Therefore, the
myopic action and the even appearances of the i.i.d.\ $\{X_\tau\}$ will
eventually make both $T_1(t)$ and $T_2(t)$ grow linearly with the elapsed time
$t$, and the forced sampling should occur only rarely. This situation differs
significantly from the traditional bandits, where the forced sampling will
inevitably make the $T_{inf}(t)$ of the order of $\sqrt{t}$, which is an
undesired result.

A detailed proof of the boundedness of ${\mathsf E}_{C_0}\{T_{inf}(t)\}$ for
this scheme is provided in \Append{\ref{app: scheme-depends-on-x}}.

%%%%%%%%%%%%%%%%%%%%%%%%%%%%%%%%%%%%%%%%%%%%%%%%%%%%%%%%%%%%%%%%%%%%%%%%%%%%%%%%%%%%%%%%%%
%
%   Wait until good
%
%%%%%%%%%%%%%%%%%%%%%%%%%%%%%%%%%%%%%%%%%%%%%%%%%%%%%%%%%%%%%%%%%%%%%%%%%%%%%%%%%%%%%%%%%%

\section{Best Arm Is Not A Function Of $X_t$\label{sec: X-depends-on-x}}
%%%%%%%%%%%%%%%%%%%%%%%%%%%%%%%%%%%%%%%%%%%%%%%%%%%%%%%%%%%%%%%%%%%%%%%%%%%%%%%%%%%%%%%%%%
%
%   Wait until good: formulations
%
%%%%%%%%%%%%%%%%%%%%%%%%%%%%%%%%%%%%%%%%%%%%%%%%%%%%%%%%%%%%%%%%%%%%%%%%%%%%%%%%%%%%%%%%%%
\subsection{Formulation}
\begin{figure}[t]\centering
\includegraphics[width=2in, keepaspectratio=true]{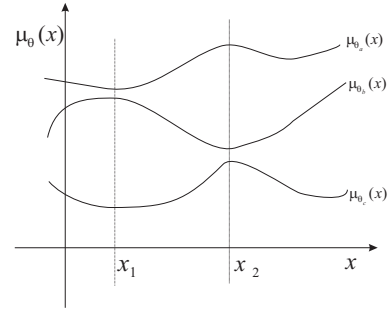}
\caption{The best arm at time $t$ {\em never} depends on the side observation
$X_t$. That is, for any possible pair, $(\theta_1,\theta_2)$, the two curves,
$\mu_{\theta_1}(x)$ and $\mu_{\theta_2}(x)$, do not intersect each other.
However, in this case, we can postpone our sampling to the most informative
time instants.
%For example, if
%$(\theta_1,\theta_2)=(\theta_a,\theta_b)$, we only perform forced
%sampling on arm 2 when $X_t=x_2$, which yields the largest
%information distance $I(\theta_b,\theta_a|x)$ as a function of
%$x$.
} \label{fig2}
\end{figure}

Besides the assumption of constant $G$, in this section, we consider the case
in which for all $C\in{\mathbf \Theta}^2$, $M_{C}(x)$ is not a function of $x$,
and we thus can use $M_C:=M_C(x)$ as shorthand notation. \Figure{\ref{fig2}}
illustrates this situation.
%\begin{description}
%\item [Independence:] $G_{C_0}=G$ does not depend on $C_0$.
%\item [Best arm as a function  of $X_t$:] $\forall C=(\theta_1,\theta_2)$, $\theta_1\neq\theta_2$,  we have either $P(\mu_{1(C)}(X_t)<\mu_{2(C)}(X_t))=1$ or $P(\mu_{1(C)}(X_t)>\mu_{2(C)}(X_t))=1$.
%\end{description}

The needed regularity conditions are similar to those in Section~\ref{sec:
depends-on-x}:
\begin{enumerate}
\item ${\mathbf X}$ is a finite set  and ${\mathsf P}_G(X_t=x)>0$ for all $x\in{\mathbf X}$.
\item $\forall \theta_1,\theta_2,x$, $I(\theta_1,\theta_2|x)$ is strictly positive  and finite.
\end{enumerate}

In this case, one arm is always better than the other no matter what value of
$X_t$ occurs. The conflict between learning and control still exists. As
expected, the growth rate of the expected inferior sampling time is again lower
bounded by $\log t$, but with the additional help of $X_t$
 we can see  improvements over the traditional bandit problems.

To greatly simplify the notation, we also assume that
\begin{enumerate}
\setcounter{enumi}{3}
\item For all  $x$, the conditional expected reward $\mu_\theta(x)$
    is strictly increasing w.r.t.\ $\theta$.
\end{enumerate}
This condition gives us the notational convenience that the order of
$(\mu_{\theta_1}(x), \mu_{\theta_2}(x))$ is simply the same as the order of
$(\theta_1,\theta_2)$.

\indent Example:
\begin{itemize}
\item ${\mathbf \Theta}=(1,\infty)$, ${\mathbf X}=\{1,2,3\}$, and the
conditional reward distribution $F_{\theta}(\cdot|x)\sim{\mathcal N}(\theta
x,1)$.
\end{itemize}

%%%%%%%%%%%%%%%%%%%%%%%%%%%%%%%%%%%%%%%%%%%%%%%%%%%%%%%%%%%%%%%%%%%%%%%%%%%%%%%%%%%%%%%%%%
%
%   Wait until good: Save the intuition part to the figure.
%
%%%%%%%%%%%%%%%%%%%%%%%%%%%%%%%%%%%%%%%%%%%%%%%%%%%%%%%%%%%%%%%%%%%%%%%%%%%%%%%%%%%%%%%%%%
%The intuition in this case is as follows.  Consider the configuration is
%$(\theta_1,\theta_2)=(\theta_a,\theta_b)$ and arm 1 is always better than arm 2. We wish to sample
%arm 1 most of the time (control part), and force sample arm 2 once in a while (learning part). With
%the help of the side information $X_t$, we can postpone our forced sampling to the most informative
%times, when $X_t=x_2$. If the authentic configuration is $(\theta_b,\theta_c)$, then the proper
%timing of forced sampling is when $X_t=x_1$.
%\begin{figure}[t]
%\includegraphics[width=7.5cm, keepaspectratio=true]{wait_until_good.eps}
%\caption{the best arm does not depend on $X_t$, so we should wait to learn until those times that
%we can get the most information.} \label{fig: wait until good}
%\end{figure}

%%%%%%%%%%%%%%%%%%%%%%%%%%%%%%%%%%%%%%%%%%%%%%%%%%%%%%%%%%%%%%%%%%%%%%%%%%%%%%%%%%%%%%%%%%
%
%   Wait until good: derivation of lower bound
%
%%%%%%%%%%%%%%%%%%%%%%%%%%%%%%%%%%%%%%%%%%%%%%%%%%%%%%%%%%%%%%%%%%%%%%%%%%%%%%%%%%%%%%%%%%
\subsection{Lower Bound}
\begin{theorem}[$\log t$ Lower Bound]
Under the above assumptions, for any uniformly good rule $\{\phi_\tau\}$,
$T_{inf}(t)$ satisfies
\begin{eqnarray}
        &&\lim_{t\rightarrow\infty}{\mathsf P}_{C_0}\left(
                T_{inf}(t)\geq\frac{(1-\epsilon)\log t}{K_{C_0}}
        \right)=1,~\forall \epsilon>0,\nonumber\\
\mbox{and~~}        && \liminf_{t\rightarrow\infty}\frac{{\mathsf
E}_{C_0}\{T_{inf}(t)\}}{\log t}\geq
\frac{1}{K_{C_0}},\label{eq:lbd-X-depends-on-x}
\end{eqnarray}
where $K_{C_0}$ is a constant depending on $C_0$. If $M_{C_0}=2$, then
$T_{inf}(t)=T_1(t)$. The constant $K_{C_0}$ can be expressed as follows.
\begin{eqnarray}
K_{C_0}=\inf_{\theta:\theta>\theta_2}\sup_{x\in{\mathbf X}}\{I(\theta_1,
\theta|x)\}.\label{eq:def-KC0-wait}
\end{eqnarray}
The expression for $K_{C_0}$ for the case in which $M_{C_0}=1$ can be obtained
by symmetry. \label{thm: lbd-X-depends-on-x}
\end{theorem}

Note 1: if the decision maker is not able to access the side observation $X_t$,
the player will then face  the {\it unconditional} reward distribution $\int_x
F_{\theta_i}(dy|x)G(dx)$ rather than $F_{\theta_i}(dy|x)$. Let
$I(\theta_1,\theta_2)$ denote the Kullback-Leibler information between the {\it
unconditional} reward distributions. By the convexity of the Kullback-Leibler
information, we have
\begin{eqnarray}
\sup_x I(\theta_1,\theta|x)\geq \int_x I(\theta_1,\theta|x)G(dx)\geq
I(\theta_1,\theta).\nonumber
\end{eqnarray}
This shows that the new constant in front of $\log t$, in
(\ref{eq:def-KC0-wait}), is no larger than the corresponding constant in
(\ref{eq:def-KC0}), and the additional side information $X_t$ generally
improves the decision made in the bandit problem. As we would expect,
\Theorem{\ref{thm: lbd-X-depends-on-x}} collapses to
\Theorem{\ref{thm:basic-theorem}} when $|{\mathbf X}|=1$.

Note 2:  This situation is like having several related bandit machines, whose
reward distributions are all determined by the common configuration pair
$(\theta_1,\theta_2)$. The information obtained from one machine is also
applicable to the other machines. If arm~2 is always better than arm~1, we wish
to sample arm~2 most of the time (the control part), and force sample arm~1
once in a while (the learning part). With the help of the side information
$X_t$,  we can postpone our forced sampling (learning) to the most informative
machine $X_t=x$. As a result, the constant in the $\log t$ lower bound in
\Theorem{\ref{thm:basic-theorem}} has been further reduced to this new
$\frac{1}{K_{C_0}}$.

A detailed proof of \Theorem{\ref{thm: lbd-X-depends-on-x}} is provided in
\Append{\ref{app: lbd-X-depends-on-x}}.

%%%%%%%%%%%%%%%%%%%%%%%%%%%%%%%%%%%%%%%%%%%%%%%%%%%%%%%%%%%%%%%%%%%%%%%%%%%%%%%%%%%%%%%%%%
%
%   Wait until good: scheme that achieves the lower bound
%
%%%%%%%%%%%%%%%%%%%%%%%%%%%%%%%%%%%%%%%%%%%%%%%%%%%%%%%%%%%%%%%%%%%%%%%%%%%%%%%%%%%%%%%%%%
\subsection{Scheme Achieving the Lower Bound\label{subsec: scheme-X-depends-on-x}}
Consider the additional conditions as follows.
\begin{enumerate}
\item $\mathbf \Theta$ is finite.
\item  A saddle point for $K_{C_0}$ exists; that
is, for all $\theta_1<\theta_2$,
\begin{eqnarray}
\inf_{\theta:\theta>\theta_2}\sup_x I(\theta_1,\theta|x)=\sup_x
\inf_{\theta:\theta>\theta_2} I(\theta_1,\theta|x).\nonumber
\end{eqnarray}
\end{enumerate}

With the above conditions, we construct a $\log t$-lower-bound-achieving scheme
$\{\phi_\tau\}$, which is inspired by \cite{AgrawalTeneketzisAnantharam89a}.
The following terms and quantities are necessary in the expression of
$\{\phi_\tau\}$.
\begin{itemize}
%\item First we recall the assumption that $\theta_1>\theta_2\Leftrightarrow\forall x,  \mu_{\theta_1}(x)>\mu_{\theta_2}(x)$.
\item Denote $\hat{C}_t:=(\theta^\alpha,\theta^\beta)$. Instead of the traditional
$(\hat{\theta}_1,\hat{\theta}_2)$ representation, we use
$(\theta^\alpha,\theta^\beta)$. Based on this representation,  we are able to
derive the following useful notation:
\begin{eqnarray}
\theta^{\alpha\wedge\beta}&:=&\min(\theta^\alpha, \theta^\beta)\nonumber\\
\alpha\wedge\beta&:=&\arg\min(\theta^\alpha, \theta^\beta)\nonumber\\
\theta^{\alpha\vee\beta}&:=&\max(\theta^\alpha,\theta^\beta)\nonumber\\
\alpha\vee\beta&:=&\arg\max(\theta^\alpha,\theta^\beta).\nonumber
\end{eqnarray}
For instance, if $\theta^\alpha<\theta^\beta$,
$\mu_{\theta^\alpha}(x)=\mu_{\theta^{\alpha\wedge\beta}}(x)$;
arm~${\alpha\wedge\beta}$ represents arm $1$; $Y_t^{\alpha\vee\beta}$ is the
reward of arm~2; and $T_{inf}(t)=T_{\alpha\wedge\beta}(t)=T_1(t)$.

\item Choose an $\epsilon$ such that $0<\epsilon<\frac{1}{2}\min\left\{\rho(F_{\theta}(\cdot|x),
F_{\vartheta}(\cdot|x)):\forall x\in{\mathbf X}, \theta\neq\vartheta\in{\mathbf
\Theta}\right\}$, where $\rho$ is the Prohorov metric. The whole system is {\it
well-sampled} if there exists a unique estimate
$\hat{C}_t=(\theta^\alpha,\theta^\beta)$, such that the empirical measure
$L_i^x(t)$  falls into the $\epsilon$-neighborhood of
$F_{i(\hat{C}_t)}(\cdot|x)$,  for all $ x\in{\mathbf X}$ and $i\in\{1,2\}$.
That is
\begin{eqnarray}
\exists
\hat{C}_t,\st\rho\left(L_i^x(t),F_{i(\hat{C}_t)}(\cdot|x)\right)<\epsilon,~~\forall
x\in{\mathbf X},i\in\{1,2\}.\nonumber
\end{eqnarray}

\item For any estimate $\hat{C}_t=(\theta^\alpha,\theta^\beta)$, define the
most informative bandit according to $\hat{C}_t$ as
\begin{eqnarray}
x^*(\hat{C}_t)&:=&\arg\max_x
\inf_{\theta:\theta>\theta^{\alpha\vee\beta}}I(\theta^{\alpha\wedge\beta},\theta,x),\nonumber
\end{eqnarray}
and $\Lambda_t(\hat{C}_t, \theta)$ to be the conditional likelihood ratio
between the seemingly inferior arm~$\theta^{\alpha\wedge\beta}$ and the
competing parameter~$\theta$:
\begin{eqnarray}
\Lambda_t(\hat{C}_t, \theta):= \prod_{m=1}^{T_{\alpha\wedge\beta}^{x^*}(t)}
\frac{F_{\theta^{\alpha\wedge\beta}}\left(dY_{\tau_{x^*}(m)}^{\alpha\wedge\beta}|x^*(\hat{C}_t)\right)}
{F_{\theta}\left(dY_{\tau_{x^*}(m)}^{\alpha\wedge\beta}|x^*(\hat{C}_t)\right)},
\nonumber
\end{eqnarray}
where $\tau_{x^*}(m)$ denotes the time instant of the $m$-th pull of
arm~$\alpha\wedge\beta$ when the side observation $X_\tau=x^*(\hat{C}_\tau)$.
\item Set a total number of $|{\mathbf X}|+|{\mathbf \Theta}|^2+|{\mathbf \Theta}|^3$ counters, including $|{\mathbf X}|$ counters, named ``\ctr{($x$)}";
 $|{\mathbf \Theta}|^2$
counters, named ``\ctr{($\hat{C}$)}" for all possible $\hat{C}\in{\mathbf
\Theta}^2$; and $|{\mathbf \Theta}|^3$ counters, named
``\ctr{($\hat{C},\theta$)}" for all possible $\hat{C}$ and $\theta$. Initially,
all counters are set to zero.
\end{itemize}
\begin{algorithm}\footnotesize
\caption{$\phi_{t+1}$, the decision at time $t+1$}\label{alg:X-depends-on-x}
\begin{algorithmic}[1]
%\STATE Set all counters, \ctr{($x$)}, \ctr{($\hat{C}$)}, and
%\ctr{($\hat{C},\theta$)},   to zero.

\IF[\dotfill\cd{0}]{there exists $i\in\{1,2\}$ and $x\in{\mathbf X}$ such that
%\IF{there exists $i\in\{1,2\}$ and $x\in{\mathbf X}$ such that
$T^x_i(t)=0$,}
    \STATE $\phi_{t+1}\leftarrow\mdt{(t+1)}$.

\ELSIF[\dotfill\cd{1}]{the whole system is not {\it well-sampled} or
$\theta^\alpha=\theta^\beta$,}\markalgline{cond1}
    \STATE $\ctr{($X_{t+1}$)}\leftarrow\ctr{($X_{t+1}$)}+1$ and
    $\phi_{t+1}\leftarrow\mdt{\ctr{($X_{t+1}$)}}$.

\ELSIF[\dotfill\cd{2}]{$\theta^{\alpha\vee\beta}=\bar{\theta}:=\max{\mathbf
\Theta}$,}\markalgline{cond2}
    \STATE $\phi_{t+1}\leftarrow1$ if it is $\theta^\alpha=\bar{\theta}$. Otherwise, $\phi_{t+1}\leftarrow2$.

\ELSE[\dotfill\cd{3}]
    \STATE   $\ctr{($\hat{C}_t$)}\leftarrow\ctr{($\hat{C}_t$)}+1$.
    \IF[\dotfill\cd{3a}]{\ctr{($\hat{C}_t$)} is odd,}\markalgline{cond3a}
        \STATE $\phi_{t+1}\leftarrow M_{\hat{C}_t}(X_{t+1})$.
    \ELSE[\dotfill\cd{3b}]\markalgline{cond3b}
        \STATE
        $\theta^*\leftarrow\arg\min\{\Lambda_t(\hat{C}_t, \theta):\theta>\theta^{\alpha\vee\beta}\}$.
        \IF[\dotfill\cd{3b1}]{$X_{t+1}= x^*(\hat{C}_t)$}
            \IF[\dotfill\cd{3b1a}]{$\Lambda_t(\hat{C}_t, \theta^*)\leq t(\log t)^2$,}
                \STATE $\ctr{($\hat{C}_t,\theta^*$)}\leftarrow\ctr{($\hat{C}_t,\theta^*$)}+1$.
                \IF[\dotfill\cd{3b1a1}]{$\exists k\in{\mathbb
                    N}\st\ctr{($\hat{C}_t,\theta^*$)}=k^2$,}\markalgline{cond3b1a1}
                    \STATE $\phi_{t+1}\leftarrow\mdt{k}$.
                \ELSE[\dotfill\cd{3b1a2}]
                    \STATE $\phi_{t+1}\leftarrow 3-M_{\hat{C}_t}(X_{t+1})$.
                \ENDIF
            \ELSE[\dotfill\cd{3b1b}]
                \STATE $\phi_{t+1}\leftarrow M_{\hat{C}_t}(X_{t+1})$.
            \ENDIF
        \ELSE[\dotfill\cd{3b2}]
            \STATE $\phi_{t+1}\leftarrow M_{\hat{C}_t}(X_{t+1})$.
        \ENDIF
    \ENDIF

 \ENDIF
\end{algorithmic}
\end{algorithm}

%The decision rule $\{\phi_{\tau}\}$ is then described in
%\Algorithm{\ref{alg:X-depends-on-x}}.

\begin{theorem}[Asymptotic Tightness]
With the above conditions, the scheme described in
\Algorithm{\ref{alg:X-depends-on-x}} achieves the $\log t $ lower bound
(\ref{eq:lbd-X-depends-on-x}), so that this $\{\phi_\tau\}$ is uniformly good
and asymptotically optimal.\label{thm: scheme-X-depends-on-x}
\end{theorem}

A complete analysis  is provided in \Append{\ref{app:
pf-scheme-X-depends-on-x}}.

%%%%%%%%%%%%%%%%%%%%%%%%%%%%%%%%%%%%%%%%%%%%%%%%%%%%%%%%%%%%%%%%%%%%%%%%%%%%%%%%%%%%%%%%%%
%
%   Combination of wait until good and expect the best
%
%%%%%%%%%%%%%%%%%%%%%%%%%%%%%%%%%%%%%%%%%%%%%%%%%%%%%%%%%%%%%%%%%%%%%%%%%%%%%%%%%%%%%%%%%%
\section{Mixed case\label{sec: mixed-case}}
%In Sections \ref{sec: depends-on-x} and \ref{sec: X-depends-on-x}, we dealt
%with the cases in which the distribution of $X_t$ is constant.
 The main
difference between Sections \ref{sec: depends-on-x} and \ref{sec:
X-depends-on-x} is that in one case, for all possible $C_0$, $X_t$ {\em always}
changes the preference order, while in the other, for all possible $C_0$, $X_t$
{\em never} changes the order. A more general case is a mixture of these two.
In this section, we consider this mixed case, which is the main result of this
paper.
%%%%%%%%%%%%%%%%%%%%%%%%%%%%%%%%%%%%%%%%%%%%%%%%%%%%%%%%%%%%%%%%%%%%%%%%%%%%%%%%%%%%%%%%%%
%
%   General case: the formulations
%
%%%%%%%%%%%%%%%%%%%%%%%%%%%%%%%%%%%%%%%%%%%%%%%%%%%%%%%%%%%%%%%%%%%%%%%%%%%%%%%%%%%%%%%%%%
\subsection{Formulation}
\begin{figure}[t]\centering
\includegraphics[width=2in, keepaspectratio=true]{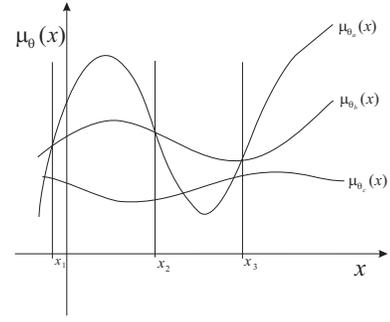}
\caption{If $(\theta_1,\theta_2)=(\theta_a, \theta_b)$, the best arm depends on
$x$, i.e.\ $\mu_{\theta_1}(x)$ and $\mu_{\theta_2}(x)$ intersect each other as
in Section \ref{sec: depends-on-x}. If
$(\theta_1,\theta_2)=(\theta_b,\theta_c)$, the best arm does not depend on $x$,
i.e.\ $\mu_{\theta_1}(x)$ and $\mu_{\theta_2}(x)$ do not intersect each other
as in Section \ref{sec: X-depends-on-x}.} \label{fig3}
\end{figure}

Besides the assumption of constant $G$, in this section, we consider the case
in which for some $C\in{\mathbf \Theta}^2$, $M_{C}(x)$ is not a function of
$x$. For the remaining $C$, there exist $x_1$ and $x_2$ s.t.\ $M_C(x_1)=1$ and
$M_{C}(x_2)=2$. For future reference, when the configuration pair $C_0$
satisfies the latter case, we say the configuration pair $C_0$ is {\it
implicitly revealing}. \Figure{\ref{fig3}} illustrates this situation.
%, which is named after the discussions in
%Section~\ref{sec: depends-on-x}.

 However, without
knowledge of the authentic underlying configuration $C_0$, we do not know
whether $C_0$ is implicitly revealing or not. In view of the results of
Sections \ref{sec: depends-on-x} and \ref{sec: X-depends-on-x}, we would like
to find a single scheme that is able to achieve bounded ${\mathsf
E}_{C_0}\{T_{inf}(t)\}$ when being applied to an implicitly revealing $C_0$,
and on the other hand to achieve the $\log t$ lower bound when being applied to
those $C_0$ which are not implicitly revealing.

The needed regularity conditions are the same as those in Sections~\ref{sec:
depends-on-x} and~\ref{sec: X-depends-on-x}:
\begin{enumerate}
\item ${\mathbf X}$ is a finite set  and ${\mathsf P}_G(X_t=x)>0$ for all $x\in{\mathbf X}$.
\item $\forall \theta_1,\theta_2,x$, $I(\theta_1,\theta_2|x)$ is strictly positive  and finite.
\end{enumerate}

To simplify the notation and the following proof, we define a partial ordering
as $\theta\prec\vartheta$ iff $\forall x,
\mu_{\theta}(x)\leq\mu_{\vartheta}(x)$, and $\theta\succ\vartheta$ is defined
similarly. Note that for a configuration $C_0=(\theta_1,\theta_2)$, it can be
the case that neither $\theta_1\prec\theta_2$ nor $\theta_1\succ\theta_2$.

\indent Example:
\begin{itemize}
\item ${\mathbf \Theta}=(0, \infty)$, ${\mathbf X}=\{-1,1\}$ and the conditional reward distribution $F_\theta(\cdot|x)\sim {\mathcal N}(\theta^2-\theta
x,1)$. Then $C_0=(\theta_1,\theta_2)=(0.1, 0.2)$ is implicitly revealing, but
$C_0=(0,10)$ is not.
\end{itemize}

%%%%%%%%%%%%%%%%%%%%%%%%%%%%%%%%%%%%%%%%%%%%%%%%%%%%%%%%%%%%%%%%%%%%%%%%%%%%%%%%%%%%%%%%%%
%
%   General case: the lower bound
%
%%%%%%%%%%%%%%%%%%%%%%%%%%%%%%%%%%%%%%%%%%%%%%%%%%%%%%%%%%%%%%%%%%%%%%%%%%%%%%%%%%%%%%%%%%
\subsection{Lower Bound}
\begin{theorem}[$\log t$ Lower Bound]
Under the above assumptions, for any uniformly good rule $\{\phi_\tau\}$, if
$C_0$ is not implicitly revealing, $T_{inf}(t)$ satisfies
\begin{eqnarray}
        &&\lim_{t\rightarrow\infty}{\mathsf P}_{C_0}\left(
                T_{inf}(t)\geq\frac{(1-\epsilon)\log t}{K_{C_0}}
        \right)=1,~\forall\epsilon>0,\nonumber\\
\mbox{and~~}        &&
\liminf_{t\rightarrow\infty}\frac{E_{C_0}\{T_{inf}(t)\}}{\log t}\geq
\frac{1}{K_{C_0}},\label{eq: lbd-mixed-case}
\end{eqnarray}
where $K_{C_0}$ is a constant depending on $C_0$. If $M_{C_0}=2$,
$T_{inf}(t)=T_1(t)$, and the constant $K_{C_0}$ can be expressed as follows.
\begin{eqnarray}
K_{C_0}=\inf_{\left\{\theta: \exists x_0,\st
\mu_{\theta}(x_0)>\mu_{\theta_2}(x_0)\right\}}\sup_x\{I(\theta_1,\theta|x)\}.\nonumber
\end{eqnarray}
The expression for $K_{C_0}$ for the case in which $M_{C_0}=1$ can be obtained
by symmetry. \label{thm: lbd-mixed-case}
\end{theorem}
The only difference between the lower bounds (\ref{eq:lbd-X-depends-on-x}) and
(\ref{eq: lbd-mixed-case}) is that, in (\ref{eq: lbd-mixed-case}), $K_{C_0}$
has been changed from taking the infimum over $\{\theta>\theta_2\}=\{\forall
x,\mu_{\theta}(x)>\mu_{\theta_2}(x) \}$ to a larger set, $\{\theta: \exists x,
\mu_{\theta}(x)>\mu_{\theta_2}(x)\}$. The reason for this is that under this
case, consider a $\theta$ for which there exists $x$ such that
$\mu_{\theta}(x)>\mu_{\theta_2}(x)$. If the authentic configuration is
$C'=(\theta,\theta_2)$ rather than $(\theta_1,\theta_2)$, a linear order of
incorrect sampling will be introduced, which violates the uniformly-good-rule
assumption. As a result, a broader class of competing distributions
$C'=(\theta,\theta_2)$ must be considered, i.e., we must consider a different
set of configurations, over which the infimum is taken.

A detailed proof is contained in \Append{\ref{app: pf-mixed-case}}.

%%%%%%%%%%%%%%%%%%%%%%%%%%%%%%%%%%%%%%%%%%%%%%%%%%%%%%%%%%%%%%%%%%%%%%%%%%%%%%%%%%%%%%%%%%
%
%   General case: intuition
%
%%%%%%%%%%%%%%%%%%%%%%%%%%%%%%%%%%%%%%%%%%%%%%%%%%%%%%%%%%%%%%%%%%%%%%%%%%%%%%%%%%%%%%%%%%
%Here we provide the intuition for the lower bound and the bound-achieving scheme.

%\noindent\textit{Intuition: } The only difference between the lower bound here and the one in
%Section \ref{sec: wait until good} is that $K_{C_0}$ has been changed from taking the infimum over
%$\{\theta\succ\theta_2\}$ to a larger set, $\{\theta:
%P\{\mu_{\theta}(X_t)>\mu_{\theta_2}(X_t)\}>0\}$. The reason is that we need to consider the
%case that $C'=(\theta,\theta_2)$, satisfying
%$P\left(\mu_{\theta}(X_t)>\mu_{\theta_2}(X_t)\right)>0$, is mis-detected as $C_0$.  To avoid having
%this mis-detection introduce a linear order of incorrect sampling, we need to do the forced
%sampling more often.  The constant in front of $\log t$ increases as a result of the relaxed
%conditions on the parameter space.

%%%%%%%%%%%%%%%%%%%%%%%%%%%%%%%%%%%%%%%%%%%%%%%%%%%%%%%%%%%%%%%%%%%%%%%%%%%%%%%%%%%%%%%%%%
%
%   General case: Scheme
%
%%%%%%%%%%%%%%%%%%%%%%%%%%%%%%%%%%%%%%%%%%%%%%%%%%%%%%%%%%%%%%%%%%%%%%%%%%%%%%%%%%%%%%%%%%
\subsection{Scheme Achieving the Lower Bound}
Consider the same two additional conditions as those in Section~\ref{sec:
X-depends-on-x}.
\begin{enumerate}
\item $\mathbf \Theta$ is finite.
\item  A saddle point for $K_{C_0}$ exists; that
is, for all $\theta_1$,
\begin{eqnarray}
&&\inf_{\left\{\theta: \exists x_0,
\mu_{\theta}(x_0)>\mu_{\theta_2}(x_0)\right\}}\sup_x
I(\theta_1,\theta|x)\nonumber\\
&&=\sup_x\inf_{\left\{\theta: \exists x_0,
\mu_{\theta}(x_0)>\mu_{\theta_2}(x_0)\right\}} I(\theta_1,\theta|x).\nonumber
\end{eqnarray}
\end{enumerate}

%With the above conditions, we construct a single scheme in
%\Algorithm{\ref{alg:mixed-case}} such that it has bounded ${\mathsf
%E}_{C_0}\{T_{inf}(t)\}$ if $C_0$ is implicitly revealing, and achieves the
%lower bound (\ref{eq: lbd-mixed-case}) when being applied to such $C_0$ that is
%not implicitly revealing.

\begin{algorithm}\footnotesize
\caption{$\phi_{t+1}$, the decision at time $t+1$}\label{alg:mixed-case}
\begin{algorithmic}[1]
%\STATE Set all counters, \ctr{($x$)}, \ctr{($\hat{C}$)}, and
%\ctr{($\hat{C},\theta$)},   to zero.

\IF[\dotfill\cd{0}]{there exists $i\in\{1,2\}$ and $x\in{\mathbf X}$ such that
$T^x_i(t)=0$,}
    \STATE $\phi_{t+1}\leftarrow\mdt{(t+1)}$.

\ELSIF[\dotfill\cd{1}]{the whole system is not {\it well-sampled} or
$\theta^\alpha=\theta^\beta$,}%\markalgline{cond1}
    \STATE $\ctr{($X_{t+1}$)}\leftarrow\ctr{($X_{t+1}$)}+1$ and
    $\phi_{t+1}\leftarrow\mdt{\ctr{($X_{t+1}$)}}$.

\ELSIF[\dotfill\cd{2}]{there exists $i\in\{1,2\}$, such that $\forall \theta,
x, \mu_\theta(x)\leq \mu_{i(\hat{C}_t)}(x)$,}\markalgline{newcond20}
    \STATE $\phi_{t+1}\leftarrow i$, where $i$ is the satisfying
    index.\markalgline{newcond21}

\ELSIF[\dotfill\cd{2.5}]{$\hat{C}_t$ is implicitly
revealing,}\markalgline{newcond250}
    \STATE $\phi_{t+1}\leftarrow M_{\hat{C}_t}(X_{t+1})$.\markalgline{newcond251}

\ELSE[\dotfill\cd{3}]
    \STATE   $\ctr{($\hat{C}_t$)}\leftarrow\ctr{($\hat{C}_t$)}+1$.
    \IF[\dotfill\cd{3a}]{\ctr{($\hat{C}_t$)} is odd,}%\markalgline{cond3a}
        \STATE $\phi_{t+1}\leftarrow M_{\hat{C}_t}(X_{t+1})$.
    \ELSE[\dotfill\cd{3b}]%\markalgline{cond3b}
        \STATE
        $\theta^*\leftarrow\arg\min\{\Lambda_t(\hat{C}_t, \theta):\forall \theta, \exists x_0, \st
        \mu_\theta(x_0)>\mu_{\theta^{\alpha\vee\beta}}(x_0)\}$.\markalgline{newcond3b1}
        \IF[\dotfill\cd{3b1}]{$X_{t+1}= x^*(\hat{C}_t)$}
            \IF[\dotfill\cd{3b1a}]{$\Lambda_t(\hat{C}_t, \theta^*)\leq t(\log t)^2$,}
                \STATE $\ctr{($\hat{C}_t,\theta^*$)}\leftarrow\ctr{($\hat{C}_t,\theta^*$)}+1$.
                \IF[\dotfill\cd{3b1a1}]{$\exists k\in{\mathbb
                    N}\st\ctr{($\hat{C}_t,\theta^*$)}=k^2$,}%\markalgline{cond3b1a1}
                    \STATE $\phi_{t+1}\leftarrow\mdt{k}$.
                \ELSE[\dotfill\cd{3b1a2}]
                    \STATE $\phi_{t+1}\leftarrow 3-M_{\hat{C}_t}(X_{t+1})$.
                \ENDIF
            \ELSE[\dotfill\cd{3b1b}]
                \STATE $\phi_{t+1}\leftarrow M_{\hat{C}_t}(X_{t+1})$.
            \ENDIF
        \ELSE[\dotfill\cd{3b2}]
            \STATE $\phi_{t+1}\leftarrow M_{\hat{C}_t}(X_{t+1})$.
        \ENDIF
    \ENDIF

 \ENDIF
\end{algorithmic}
\end{algorithm}

A proposed scheme is described in \Algorithm{\ref{alg:mixed-case}}, which is
similar to the scheme in Section~\ref{subsec: scheme-X-depends-on-x}.   The
only differences are the insertion of \cd{2.5}, Lines~\readalgline{newcond250}
and~\readalgline{newcond251}; the modification of \cd{2},
Lines~\readalgline{newcond20} and~\readalgline{newcond21}; and the modification
of \cd{3b}, line~\readalgline{newcond3b1}.

\noindent {\it Notes:}
\begin{enumerate}
\item When the estimate $\hat{C}_t=(\theta^\alpha, \theta^\beta)$ is not
implicitly revealing, an ordering between $\theta^\alpha$ and $\theta^\beta$
exists. As a result, all notation regarding $\alpha\vee\beta$,
$\theta^{\alpha\vee\beta}$, etc., remains valid.

\item The definition of $\Lambda_t(\hat{C}_t,\theta)$ is slightly different.
For any  estimate $\hat{C}_t=(\theta^\alpha,\theta^\beta)$ that is not
implicitly revealing, we can define the most informative bandit according to
$\hat{C}_t$ as {\footnotesize
\begin{eqnarray}
&&x^*(\hat{C}_t)\label{eq:new-xstar}\\
&&:=\arg\max_x \inf_{\{\theta:\exists x_0, \mu_{\theta}(x_0)
>
\mu_{\theta^{\alpha\vee\beta}}(x_0)\}}I(\theta^{\alpha\wedge\beta},\theta|x),\nonumber
\end{eqnarray}}\cmpt
and $\Lambda_t(\hat{C}_t, \theta)$ to be the conditional likelihood ratio
between the seemingly inferior arm~$\theta^{\alpha\wedge\beta}$ and the
competing parameter~$\theta$. That is,
\begin{eqnarray}
\Lambda_t(\hat{C}_t, \theta):= \prod_{m=1}^{T_{\alpha\wedge\beta}^{x^*}(t)}
\frac{F_{\theta^{\alpha\wedge\beta}}\left(\left.dY_{\tau_{x^*}(m)}^{\alpha\wedge\beta}\right|x^*(\hat{C}_t)\right)}
{F_{\theta}\left(\left.dY_{\tau_{x^*}(m)}^{\alpha\wedge\beta}\right|x^*(\hat{C}_t)\right)},
\nonumber
\end{eqnarray}
where $\tau_{x^*}(m)$ denotes the time instant of the $m$-th pull of
arm~$\alpha\wedge\beta$ when the side observation $X_\tau=x^*(\hat{C}_\tau)$.
(The difference between this new $\Lambda_t(\hat{C}_t,\theta)$ and the previous
one in \Algorithm{\ref{alg:X-depends-on-x}} is that we have a new
$x^*(\hat{C}_t)$ defined in (\ref{eq:new-xstar}).)
\end{enumerate}

\begin{theorem}[Asymptotic Tightness]
With the above conditions, the scheme described in
\Algorithm{\ref{alg:mixed-case}} has bounded $\lim_t{\mathsf E
}_{C_0}\{T_{inf}(t)\}$, or achieves the $\log t$ lower bound (\ref{eq:
lbd-mixed-case}), depending on whether the underlying configuration pair $C_0$
is implicitly revealing or not.\label{thm: scheme-mixed-case}
\end{theorem}
%The performance achieved in both Sections \ref{sec: depends-on-x} and \ref{sec:
%X-depends-on-x} has been attained here, so that
%\Algorithm{\ref{alg:mixed-case}} is asymptotically optimal.

A detailed analysis is given in \Append{\ref{app: pf-mixed-case}}.

%%%%%%%%%%%%%%%%%%%%%%%%%%%%%%%%%%%%%%%%%%%%%%%%%%%%%%%%%%%%%%%%%%%%%%%%%%%%%%%%%%%%%%%%%%
%
%   Conclusion
%
%%%%%%%%%%%%%%%%%%%%%%%%%%%%%%%%%%%%%%%%%%%%%%%%%%%%%%%%%%%%%%%%%%%%%%%%%%%%%%%%%%%%%%%%%
\section{Conclusion}
\def\tlc#1{\parbox[c]{4.5cm}{\vspace{.1cm}#1}}
\def\tcc#1{\parbox[c]{5.3cm}{\vspace{.1cm}#1}}
\def\trc#1{\parbox[c]{6.5cm}{\vspace{.1cm}#1}}

\begin{table*}[t]
 \caption{Summary of the Benefit of the Side Observations and the Required Regularity Conditions.}
    \label{tab:summary}
\begin{center}{\footnotesize
    \begin{tabular}{ccc}
        \hline
        \hline
        \tlc{Characterization} & \tcc{Regularity Conditions}  & \trc{Results}\\
        \hline
            \tlc{$G_{C_1}\neq
                G_{C_2}$ iff $C_1\neq C_2$.}
            & \tcc{ As $\hat{C}_t\rightarrow C_0$, $\forall x$, $M_{\hat{C}_t}(x)=M_{C_0}(x)$.}
            &\trc{$\exists \{\phi_\tau\}$ \st $\forall C_0$, $\lim_t{\mathsf E}_{C_0}\{T_{inf}(t)\}<\infty$.}\\
        \hline
            \tlc{(i) Constant $G_C$, i.e., $G_C:=G$,\\
                (ii) $\forall C, \exists x_1,x_2$, \st $M_C(x_1)=1$,\\ $M_C(x_2)=2$ (implicitly revealing).}
            & \tcc{(i) $\mathbf X$ is finite.\\
                                        (ii) $\forall \theta_1\neq\theta_2,x$, $0<I(\theta_1,\theta_2|x)<\infty$.\\
                                   (iii) $\forall x$, $\mu_\theta(x)$ is continuous w.r.t.\ $\theta$.}
            &\trc{$\exists \{\phi_\tau\}$ such that $\forall C_0$, $\lim_t{\mathsf E}_{C_0}\{T_{inf}(t)\}<\infty$.}\\
        \hline
            \tlc{(i) Constant $G_C$, i.e., $G_C:=G$,\\
                                    (ii) $\forall C$, $M_C(x)$ only depends on $C$, not on $x$.
                                    }
            & \tcc{(i) $\mathbf X$ is finite.\\
                                     (ii) $\forall \theta_1\neq\theta_2,x$,
                                     $0<I(\theta_1,\theta_2|x)<\infty$,\\
                                    (iii) $\forall x$, $\mu_\theta(x)$ is strictly increasing w.r.t.\
                                    $\theta$.
                                     }
            & \multicolumn{1}{c}{
            \begin{tabular}{c}
            \trc{ For any  uniformly good $\{\phi_\tau\}$, we have $\lim_t\frac{{\mathsf E}_{C_0}\{T_{inf}(t)\}}{\log
            t}\geq\frac{1}{K_{C_0}}$,\\
                 $K_{C_0}:=\inf_\theta\sup_xI(\theta_1,\theta|x)$.}\\
            \hline
            \trc{ For finite $\mathbf \Theta$, $\exists \{\phi_\tau\}$, \st
            $\lim_t\frac{{\mathsf E}_{C_0}\{T_{inf}(t)\}}{\log
            t}\leq\frac{1}{K_{C_0}}$.}
            \end{tabular}
            }\\
        \hline
            \tlc{(i)  Constant $G_C$, i.e., $G_C:=G$,\\
                                    (ii) The underlying $
                                    C_0$ may be implicitly revealing or not.}
            & \tcc{(i) $\mathbf X$ is finite.\\
                                     (ii) $\forall \theta_1\neq\theta_2,x$,
                                     $0<I(\theta_1,\theta_2|x)<\infty$.
                                     }
            & \multicolumn{1}{c}{
            \begin{tabular}{c}
            \trc{For any  uniformly good $\{\phi_\tau\}$, if $C_0$ is not implicitly revealing,
                        we have $\lim_t\frac{{\mathsf E}_{C_0}\{T_{inf}(t)\}}{\log t}\geq\frac{1}{K_{C_0}}$,
                        $K_{C_0}:=\inf_\theta\sup_xI(\theta_1,\theta|x)$.}\\
            \hline
            \trc{For finite $\mathbf \Theta$, $\exists \{\phi_t\}$ \st\\ (1)~if $C_0$ is implicitly revealing,
                        ${\mathsf
            E}_{C_0}\{T_{inf}(t)\}<\infty$,\\
                 (2)~if $C_0$ is not {\it i.r.}, $\lim_t\frac{{\mathsf E}_{C_0}\{T_{inf}(t)\}}{\log t}\leq\frac{1}{K_{C_0}}$.}
            \end{tabular}
            }\\
        \hline
        \hline
    \end{tabular}
    }
\end{center}
\end{table*}

We have shown that observing additional side information can significantly
improve sequential decisions in bandit problems. If the side observation itself
directly provides information about the underlying configuration, then it
resolves the dilemma of forced sampling and optimal control. The expected
inferior sampling time will be bounded, as shown in Section \ref{sec: d-info}.
If the side observation does not provide information on the underlying
configuration $(\theta_1,\theta_2)$, but {\em always} affects the preference
order (implicitly revealing), then the myopic approach of sampling the
seemingly-best arm will automatically sample both arms enough. The expected
inferior sampling time is bounded, as shown in Section \ref{sec: depends-on-x}.
If the side observation {\em does not} affect the preference order at all, the
dilemma still exists. However, by postponing our forced sampling to the most
informative time instants, we can reduce the constant in the $\log t$ lower
bound, as shown in Section \ref{sec: X-depends-on-x}.  In Section \ref{sec:
mixed-case}, we combined the settings of Sections \ref{sec: depends-on-x} and
\ref{sec: X-depends-on-x}, and obtained a general result. When the underlying
configuration $C_0$ is implicitly revealing (such that $X_t$ will change the
preference order), we obtain bounded expected inferior sampling time as in
Section \ref{sec: depends-on-x}. Even if $C_0$ is not implicitly revealing (in
that $X_t$ does not change the preference order), the new $\log t$ lower bound
can be achieved as in Section \ref{sec: X-depends-on-x}. Our results are
summarized in \Table{\ref{tab:summary}}.

% if have a single appendix:
%\appendix[Proof of the Zonklar Equations]
% or
%\appendix  % for no appendix heading
% do not use \section anymore after \appendix, only \section*
% is possibly needed

% use appendices with more than one appendix
% then use \section to start each appendix
% you must declare a \section before using any
% \subsection or using \label (\appendices by itself
% starts a section numbered zero.)
%
% Use this command to get the appendices' numbers in "A", "B" instead of the
% default capitalized Roman numerals ("I", "II", etc.).
% However, the capital letter form may result in awkward subsection numbers
% (such as "A-A"). Capitalized Roman numerals are the default.
%\useRomanappendicesfalse
%
\appendices
\section{Sanov's theorem and the Prohorov metric\label{subsec:sanov-prohorov}}
For two distributions $P$ and $Q$ on the reals, the Prohorov metric is defined
as follows.
\begin{definition}[The Prohorov metric]
For any closed set $A\subset {\mathbb R}$ and $\epsilon>0$, define
$A^\epsilon$, the $\epsilon$-flattening of $A$, as
\begin{eqnarray}
A^\epsilon:=\left\{x\in{\mathbb R}:\inf_{y\in
A}|x-y|<\epsilon\right\}.\nonumber
\end{eqnarray}
The Prohorov metric $\rho$ is then defined as follows.
\begin{eqnarray}
\rho(P,Q):=\inf\left\{\epsilon>0:P(A)\leq Q(A^\epsilon)+\epsilon,\right.
\nonumber\\
\left.\mbox{ for all closed $A\subset{\mathbb R}$.}\right\}.\nonumber
\end{eqnarray}
\end{definition}
The Prohorov metric generates the topology corresponding to convergence in
distribution. Throughout this paper, the open/closed sets on the space of
distributions are thus defined accordingly.
\begin{theorem}[Sanov's theorem]\label{thm:Sanov}
Let $L_X(n)$ denote the empirical measure of the real-valued i.i.d.\ random
variables $X_1,X_2,\cdots, X_n$. Suppose $X_i$ is of distribution $P$ and
consider any open set $A$ and closed set $B$ from the topological space of
distributions, generated by the Prohorov metric. We have
\begin{eqnarray}
\liminf_{n\rightarrow\infty}\frac{1}{n}\log {\mathsf P}_P(L_X(n)\in A)&\geq&
-\inf_{Q\in
A}I(Q,P)\nonumber\\
\limsup_{n\rightarrow\infty}\frac{1}{n}\log {\mathsf P}_P(L_X(n)\in B)&\leq&
-\inf_{Q\in B}I(Q,P).\nonumber
\end{eqnarray}
%where $I(Q,P):={\mathsf E}_{Q}\log\frac{dQ}{dP}$ is the Kullback-Leibler
%information number.
\end{theorem}
%The above is not the most general form of Sanov's theorem, but is sufficient
%for this paper.
Further discussion of the Prohorov metric and Sanov's theorem
can be found in \cite{Bucklew90,DemboZeitouni90}.

%%%%%%%%%%%%%%%%%%%%%%%%%%%%%%%%%%%%%%%%%%%%%%%%%%%%%%%%%%%%%%%%%%%%%%%%%%%%%%%%%%%%%%%%%%
%
%   Direct information: bounded scheme, Proof
%
%%%%%%%%%%%%%%%%%%%%%%%%%%%%%%%%%%%%%%%%%%%%%%%%%%%%%%%%%%%%%%%%%%%%%%%%%%%%%%%%%%%%%%%%%%
\section{Proof of \Theorem{\ref{thm: d-info}} \label{app: pf-d-info}}

\begin{proof} For any underlying  configuration pair
$C_0=(\theta_1, \theta_2)$, define the error  set ${\mathbf C}_e$ as follows.
\begin{eqnarray}
   {\mathbf C}_e&:=&\bigcup_{x\in{\mathbf X}}\{ C\in{\mathbf
\Theta}^2:
        M_C(x)\neq M_{C_0}(x)
    \}.\label{eq: error-set}
\end{eqnarray}
Let $\bar{{\mathbf C}}_e$ denote the closure of ${\mathbf C}_e$. By
\Condition{\ref{cond: d-info}}, $C_0\notin\bar{{\mathbf C}}_e$. For any $t$, we
can write {\footnotesize
\begin{eqnarray}
        {\mathsf P}_{C_0}\left(\phi_t\neq M_{C_0}(X_t)\right)&=&{\mathsf P}_{C_0}\left(M_{\hat{C}_{t}}(X_t)\neq M_{C_0}(X_t)\right)\nonumber\\
        &\leq&{\mathsf P}_{C_0}\left(\exists x, M_{\hat{C}_{t}}(x)\neq M_{C_0}(x)\right)\nonumber\\
        &=&{\mathsf P}_{C_0}\left(\hat{C}_{t}\in{\mathbf C}_e\right)\nonumber\\
        &\leq&
        {\mathsf P}_{C_0}\left(\hat{C}_{t}\in\bar{{\mathbf C}}_e\right).\nonumber
\end{eqnarray}}\cmpt
Let $\epsilon=\frac{1}{3}\inf\{\rho(G_{C_0},G_{C_e}):C_e\in\bar{{\mathbf
C}}_e\}$, which is strictly positive by \Condition{\ref{cond: d-info}}, and
consider sufficiently large $t\geq \frac{1}{\epsilon}$. If $\rho(G_{C_0},
L_X(t))<\epsilon$, then by the definition of ${\mathbf C}_t$,
$\rho(G_{\hat{C}_t},L_X(t))<\epsilon+\epsilon=2\epsilon$. By the triangle
inequality, $\rho(G_{C_0}, G_{\hat{C}_t})<3\epsilon$ and $\hat{C}_t\neq
\bar{{\mathbf C}}_e$. As a result,
\begin{eqnarray}
        \{\hat{C}_t\in\bar{{\mathbf C}}_e\}\subset\left\{\rho(L_X(t),
        G_{C_0})\geq\epsilon\right\}\stackrel{\Delta}{=}{\mathbf K}_t.\nonumber
\end{eqnarray}
${\mathbf K}_t$ is a closed set. By Sanov's theorem, the probability of
${\mathbf K}_t$ is exponentially upper bounded w.r.t.\ $t$, and so is ${\mathsf
P}_{C_0}\left(\hat{C}_{t}\in\bar{{\mathbf C}}_e\right)$. As a result, we have
\begin{eqnarray}
\lim_{t\rightarrow\infty}{\mathsf
E}_{C_0}\{T_{inf}(t)\}=\lim_{t\rightarrow\infty}\sum_{\tau=1}^t{\mathsf
P}_{C_0}(\phi_\tau\neq M_{C_0}(X_\tau))<\infty.\nonumber
\end{eqnarray}
By the monotone convergence theorem,
% we also have
%\begin{eqnarray}
%\lim_{t\rightarrow\infty}{\mathsf E}_{C_0}\{T_{inf}(t)\}={\mathsf E}_{C_0}
%\{\lim_{t\rightarrow\infty}T_{inf}(t)\}<\infty.\nonumber
%\end{eqnarray}
%Since
the expectation of $\lim_{t\rightarrow\infty}T_{inf}(t)$ is finite, which
implies that $\lim_{t\rightarrow\infty}T_{inf}(t)$ is finite~a.s.
\end{proof}

%%%%%%%%%%%%%%%%%%%%%%%%%%%%%%%%%%%%%%%%%%%%%%%%%%%%%%%%%%%%%%%%%%%%%%%%%%%%%%%%%%%%%%%%%%
%
%   The analysis of theorem expect the best
%
%%%%%%%%%%%%%%%%%%%%%%%%%%%%%%%%%%%%%%%%%%%%%%%%%%%%%%%%%%%%%%%%%%%%%%%%%%%%%%%%%%%%%%%%%%
\section{Proof of \Theorem{\ref{thm: depends-on-x}}\label{app: scheme-depends-on-x}}
%%%%%%%%%%%%%%%%%%%%%%%%%%%%%%%%%%%%%%%%%%%%%%%%%%%%%%%%%%%%%%%%%%%%%%%%%%%%%%%%%%%%%%%%%%
%
%   Expecting the best: necessary lemma
%
%%%%%%%%%%%%%%%%%%%%%%%%%%%%%%%%%%%%%%%%%%%%%%%%%%%%%%%%%%%%%%%%%%%%%%%%%%%%%%%%%%%%%%%%%%
Similarly, we define ${\mathbf C}_e$ as that in (\ref{eq: error-set}). We need
the following lemma to complete the analysis.
\begin{lemma} With the regularity conditions specified
in Section \ref{sec: depends-on-x}, $\exists a_1, a_2>0$ such that ${\mathsf
P}_{C_0}(\hat{C}_t\in{\mathbf C}_e)\leq
a_1\exp\left(-a_2\min\left\{T_1^{x^\star}(t),T_2^{x^\star}(t)\right\}\right)$.\label{lem:
1}
\end{lemma}

\begin{thmproof}{Proof of \Lemma{\ref{lem: 1}}:} By the continuity of $\mu_\theta(x)$ w.r.t.\
$\theta$ and the assumption of finite $\mathbf X$, it can be shown that
$C_0\in\bar{\mathbf C}_e^c$.\footnote{$\bar{\mathbf C}_e^c$ denotes the
complement of $\bar{{\mathbf C}}_e$.} Therefore there exists a neighborhood of
$C_0$, ${\mathbf
C}_{\delta}=(\theta_1-\delta,\theta_1+\delta)\times(\theta_2-\delta,\theta_2+\delta)$,
such that ${\mathbf C}_\delta\subset\bar{{\mathbf
C}}_e^c\Leftrightarrow\bar{{\mathbf C}}_e\subset{\mathbf C}_\delta^c$.

Define a strictly positive $\epsilon>0$ as follows.
\begin{eqnarray}
        \epsilon:=\frac{1}{4}\inf\left\{
                \rho\left(F_{\hat{\theta}_i}(\cdot|x),F_{\theta_i}(\cdot|x)\right):~~~~~~~~~~~~~~~\right.\nonumber\\
                \left.\forall x\in{\mathbf X}, i\in\{1,2\},
                (\hat{\theta}_1,\hat{\theta}_2)\in{\mathbf C}_\delta^c
        \right\}.\nonumber
\end{eqnarray}
We would like to prove that for sufficiently  large $t>\frac{1}{\epsilon}$,
\begin{eqnarray}
\left\{\hat{C}_t\in{\mathbf C}_\delta^c\right\}\subset\left\{\exists i,
\rho\left(L_i^{x^\star}(t),
F_{\theta_i}(\cdot|x^\star)\right)>\epsilon\right\}.\nonumber
\end{eqnarray}
%We consider the case when the empirical measure $L_i^{x^\star}(t)$ is close to
%the true distribution $F_{\theta_i}(\cdot|x^\star)$ and use proof by
%contradiction.
Suppose $\rho\left(L_i^{x^\star}(t),
F_{\theta_i}(\cdot|x^\star)\right)\leq\epsilon$ for both $i=1,2$. By the
definition of $\sigma(C_0,t)$, we have
\begin{eqnarray}
    \sigma(C_0,t)\leq2\epsilon.\label{eq:C-2epsilon}
\end{eqnarray}
However, for those $\hat{C}_t\in{\mathbf C}_\delta^c$,
 by the definition of $\epsilon$, for some $i\in\{1,2\}$, we
have
\begin{eqnarray}
\sigma(\hat{C}_t,t)
&\geq& \rho(F_{\hat{\theta}_i}(\cdot|x^\star), L_i^{x^\star}(t))\nonumber\\
&\geq& \rho(F_{\hat{\theta}_i}(\cdot|x^\star),
F_{\theta_i}(\cdot|x^\star))-\rho(F_{\theta_i}(\cdot|x^\star),
L_i^{x^\star}(t))\nonumber\\
&\geq&3\epsilon,\label{eq:C-3epsilon}
\end{eqnarray}
which contradicts the definition of ${\mathbf C}_t$ since (\ref{eq:C-2epsilon})
and (\ref{eq:C-3epsilon}) imply
$\sigma(\hat{C}_t,t)>\frac{1}{t}+\sigma(C_0,t)$. As a result, for sufficiently
large $t$, we have
\begin{eqnarray}
        \left\{\hat{C}_t\in{\mathbf C}_e\right\}&\subset&\left\{\hat{C}_t\in{\mathbf C}_\delta^c\right\}\nonumber\\
        &\subset&\left\{\exists i, \rho\left(L_i^{x^\star}(t), F_{\theta_i}(\cdot|x^\star)\right)>\epsilon\right\}\nonumber\\
        &=&\bigcup_{i=1,2}\left\{ \rho\left(L_i^{x^\star}(t), F_{\theta_i}(\cdot|x^\star)\right)>\epsilon\right\}.\label{eq: Subterm1OfExpctTheBest}
\end{eqnarray}
By Sanov's theorem, the probability of each term in the union of the right-hand
side of (\ref{eq: Subterm1OfExpctTheBest}) is exponentially bounded w.r.t.\
$T_i^{x^\star}(t)$. As a result, the probability of this finite union is
bounded by $a_1\exp\left(-a_2\min\left\{T_1^{x^\star}(t),
T_2^{x^\star}(t)\right\}\right)$ for some $a_1$, $a_2>0$.
\end{thmproof}

%%%%%%%%%%%%%%%%%%%%%%%%%%%%%%%%%%%%%%%%%%%%%%%%%%%%%%%%%%%%%%%%%%%%%%%%%%%%%%%%%%%%%%%%%%
%
%   Expecting the best: show that its expectation is bounded
%
%%%%%%%%%%%%%%%%%%%%%%%%%%%%%%%%%%%%%%%%%%%%%%%%%%%%%%%%%%%%%%%%%%%%%%%%%%%%%%%%%%%%%%%%%%
\begin{thmproof}{Analysis of the scheme:}
We first use induction to show that $\forall t\geq 6$, $T_i(t)\geq \sqrt{t}$.
This statement is true for $t=6$. Suppose $T_t(t-1)\geq \sqrt{t-1}$. If
$T_i(t-1)\geq \sqrt{t}$, by the monotonicity of $T_i(t)$ w.r.t.\ $t$, we have
$T_i(t)\geq T_i(t-1)\geq \sqrt{t}$. If $T_i(t-1)< \sqrt{t}$, by the forced
sampling mechanism, $T_i(t)=T_i(t-1)+1\geq \sqrt{t-1}+1\geq \sqrt{t}$.

We consider the event of the inferior sampling at time $(t+1)$:{\footnotesize
\begin{eqnarray}
        \left\{\phi_{t+1}\neq M_{C_0}(X_{t+1})\right\}%\nonumber\\
        &=&\left\{\phi_{t+1}\neq M_{C_0}(X_{t+1}),\hat{C}_t\in{\mathbf C}_e\right\}\nonumber\\
        &&~~\cup\left\{\phi_{t+1}\neq M_{C_0}(X_{t+1}),\hat{C}_t\in{\mathbf C}_e^c\right\}\nonumber\\
        &\subset&\left\{\hat{C}_t\in{\mathbf C}_e\right\}\nonumber\\
        &&~~\cup\left\{\phi_{t+1}\neq M_{C_0}(X_{t+1}), \hat{C}_t\in{\mathbf C}_e^c\right\}\nonumber\\
        &\stackrel{\Delta}{=}&A_{t+1}\cup B_{t+1}.\label{eq: A-B}
\end{eqnarray}}\cmpt
Since $T_i(t)\geq \sqrt{t}$, we have $\min_i T_i(t)\geq\sqrt{t}$ and $\min_i
T_i^{x^\star}(t)\geq\frac{\sqrt{t}}{|{\mathbf X}|}$. By \Lemma{\ref{lem: 1}},
we have ${\mathsf P}_{C_0}(A_{t+1})\leq a_1e^{-a_2\frac{\sqrt{t}}{|{\mathbf
X}|}}$, and hence $\sum^\infty_{t+1=7}{\mathsf P}_{C_0}(A_{t+1})<\infty$.

For ${\mathsf P}_{C_0}(B_{t+1})$,  we can write{\footnotesize
\begin{eqnarray}
        B_{t+1}&=&\left\{\phi_{t+1}\neq M_{\hat{C}_t}(X_{t+1}),
                \hat{C}_t\in{\mathbf C}_e^c\right\}\nonumber\\
        &\subset&\left\{\min\{ T_i(t)\}_i<\sqrt{t+1},\hat{C}_t\in{\mathbf C}_e^c\right\}\nonumber\\
        &=&\left\{\min\{ T_i(t)\}_i=\sqrt{t}\in {\mathbb N},\hat{C}_t\in{\mathbf C}_e^c\right\}\nonumber\\
        &\subset&\left\{\exists i,\phi_a\neq i,\forall a\in(\tau_0,t],T_i(\tau_0)=\sqrt{t}\right\}\nonumber\\
        &\stackrel{\Delta}{=}&B_{t+1}^1\cup B_{t+1}^2,\label{eqn3}
\end{eqnarray}}\cmpt
where $\tau_0=\left(\sqrt{t}-1\right)^2+1$ and $B_{t+1}^1$, $B_{t+1}^2$
correspond to $i=1,2$, respectively. The first equality comes from the fact
that since $\hat{C}_t\in{\mathbf C}_e^c$,
$M_{C_0}(X_{t+1})=M_{\hat{C}_t}(X_{t+1})$. The first subset sign comes from the
fact that $\phi_{t+1}\neq M_{\hat{C}_t}(X_{t+1})$ implies the decision rule
$\phi_{t+1}$ is in the stage of forced sampling. The second equality follows by
combining both the inequalities: $\min \{T_i(t)\}_i\geq\sqrt{t}$ and $\min
\{T_i(t)\}_i<\sqrt{t+1}$ and the fact that both $t$ and $T_i(t)$ are integers.

The reasoning behind the second subset inequality is as follows. By again using
the fact that $T_i(t)\geq\sqrt{t}$ and substituting $\tau_0$ for $t$, we have
$\sqrt{t}-1<T_i(\tau_0)$ and thus have $T_i(\tau_0)=\sqrt{t}=T_i(t)$, which
guarantees that arm~$i$ has not been sampled from time~$\tau_0+1$ to~$t$.

By the symmetry between $B_{t+1}^1$ and $B_{t+1}^2$, we can consider only
$B_{t+1}^1$ for example. We have{\footnotesize
\begin{eqnarray}
        &&{\mathsf P}_{C_0}\left(B_{t+1}^1\right)\nonumber\\
        &&\leq~{\mathsf P}_{C_0}\left(M_{\hat{C}_{a-1}}(X_a)=2,\forall a\in(\tau_0,t]\right)\nonumber\\
        &&=\prod_{a\in(\tau_0,t]}{\mathsf  P}_{C_0}\left(
            \left.M_{\hat{C}_{a-1}}(X_a)=2\right|M_{\hat{C}_{b-1}}(X_b)=2,\forall b\in(\tau_0,a)\right)\nonumber\\
        &&\leq~ \left(1-\min_x\{{\mathsf P}_{G}(X_t=x)\}\right)^{t-\tau_0}.\label{eqn1}
\end{eqnarray}}\cmpt
The first inequality comes from the definition of $\{\phi_\tau\}$ which implies
that if $T_1(\tau_0)=T_1(t)\geq\sqrt{t}$, the forced sampling mechanism is not
active during the time interval $(\tau_0,t]$. So $\phi_a=2$ implies
$M_{\hat{C}_{a-1}}(X_a)=2$, $\forall a\in(\tau_0,t]$. The  second inequality
comes from the assumption of i.i.d.\ $\{X_\tau\}$, which implies that $X_a$ is
independent of $\hat{C}_{b}$ and $X_b$ for all $b< a$. Since at least one $x$
will make $M_{\hat{C}_{a-1}}(X_a)=1$, each term in the product is then upper
bounded by $1-\min_x\{{\mathsf P}_{G}(X_t=x)\}$. It is worth noting that by the
regularity assumption on $G$,  $1-\min_x\{{\mathsf P}_{G}(X_t=x)\}$ is strictly
less than $1$.

Then from  (\ref{eqn3}), (\ref{eqn1}), and the union bound, we obtain ${\mathsf
P}_{C_0}(B_{t+1})\leq {\mathsf P}_{C_0}(B_{t+1}^1)+{\mathsf
P}_{C_0}(B_{t+1}^2)\leq 2a^{t-((\sqrt{t}-1)^2+1)}$ for some $a<1$. Hence
$\sum_{t+1=7}^\infty {\mathsf P}_{C_0}(B_{t+1})<\infty$. From (\ref{eq: A-B}),
we conclude that
\begin{eqnarray}
&&\lim_{t\rightarrow\infty}{\mathsf
E}_{C_0}\left\{T_{inf}(t)\right\}\nonumber\\
&&\leq~6+\sum_{\tau+1=7}^\infty ({\mathsf P}_{C_0}(A_{\tau+1})+{\mathsf
P}_{C_0}(B_{\tau+1}))<\infty,\nonumber
\end{eqnarray}
which completes the proof.
\end{thmproof}

%%%%%%%%%%%%%%%%%%%%%%%%%%%%%%%%%%%%%%%%%%%%%%%%%%%%%%%%%%%%%%%%%%%%%%%%%%%%%%%%%%%%%%%%%%
%
%   Proof of the lower bound of wait until good
%
%%%%%%%%%%%%%%%%%%%%%%%%%%%%%%%%%%%%%%%%%%%%%%%%%%%%%%%%%%%%%%%%%%%%%%%%%%%%%%%%%%%%%%%%%%
\section{Proof of \Theorem{\ref{thm: lbd-X-depends-on-x}}\label{app: lbd-X-depends-on-x}}
\begin{proof} The proof is inspired by \cite{AnantharamVaraiyaWalrand87a}. Without loss of generality, we assume
$M_{C_0}=2$, which immediately implies $T_{inf}(t)=T_1(t)$. Fix a $\theta$ with
$\mu_\theta>\mu_{\theta_2}$, and define $C'=(\theta,\theta_2)$.
% This proof is
%mainly based on a change-of-measure argument such that if under the
%distribution ${\mathsf P}_{C_0}$, the probability of the event that the
%inferior sampling time falls below the $\log t$ lower bound is bounded away
%from zero, then this decision rule $\{\phi_\tau\}$ will not be uniformly good
%under the competing distribution ${\mathsf P}_{C'}$.
Let $\lambda(n)$ denote the log likelihood ratio between $\theta_1$ and
$\theta$ based on the first $n$ observed rewards of arm~1. That is
\begin{eqnarray}
        \lambda(n):=\sum^n_{m=1}\log\left(
                \frac{F_{\theta_1}(dY^1_{\tau(m)}|X_{\tau(m)})}{F_\theta(dY^1_{\tau(m)}|X_{\tau(m)})}
        \right),\nonumber
\end{eqnarray}
where $\tau(m)$ is a random variable corresponding to the time index of the
$m$-th pull of arm~1.

By conditioning on the sequence $\{X_{\tau(m)}\}$, $\lambda(n)$ is a sum of
independent r.v.'s. %, i.e., $\log\left(
%                \frac{F_{\theta_1}(dY^1_{\tau(m)}|X_{\tau(m)})}{F_\theta(dY^1_{\tau(m)}|X_{\tau(m)})}\right)$.
Let $K_{C'}:=\sup_{x\in{\mathbf X}}\{ I(\theta_1,\theta|x)\}$, and suppose
there exists $\delta>0$ such that
\begin{eqnarray}
        \limsup_{n\rightarrow\infty} \frac{\lambda(n)}{n}> K_{C'}+\delta,\nonumber
\end{eqnarray}
with positive probability. Then with positive probability, there exists an
$x_0$ such that the average of the subsequence for which $X_{\tau(m)}=x_0$,
will be larger than $K_{C'}+\delta$. This, however, contradicts the strong law
of large numbers since the subsequence is i.i.d.\ and with marginal expectation
$I(\theta_1,\theta|x_0)$. Thus we obtain
\begin{eqnarray}
        \limsup_{n\rightarrow\infty} \frac{\lambda(n)}{n}\leq K_{C'},~~{\mathsf P}_{C_0}-a.s.\label{eq:proof-the-max-conv1}
\end{eqnarray}
The inequality~(\ref{eq:proof-the-max-conv1}) is equivalent to the statement
that with probability one, there are finitely many $n$ such that
$\lambda(n)>n(K_{C'}+\delta)$ for some $\delta>0$. And since $K_{C'}>0$, this
in turn implies there are at most finitely manly  $n$ such that $\max_{m\leq
n}\lambda(m)>n(K_{C'}+\delta)$. As a result,  we have
\begin{eqnarray}
        \limsup_{n\rightarrow\infty}\frac{\max_{m\leq n}\lambda(m)}{n}\leq K_{C'},~~P_{C_0}-a.s.,\nonumber
\end{eqnarray}
\begin{eqnarray}
\mbox{and~~~}\lim_{n\rightarrow\infty}  {\mathsf P}_{C_0}\left(\exists m\leq n,
\lambda(m)\geq (1+\delta)n K_{C'}\right)=0.~~~ \label{eq:thm5-1}
\end{eqnarray}

Henceforth, we proceed using contradiction. Suppose
\begin{eqnarray}
\limsup_{t\rightarrow\infty} {\mathsf P}_{C_0}\left(T_1(t)<\frac{\log
t}{(1+2\delta)K_{C'}}\right)>0.\nonumber
\end{eqnarray}
Using $A_1$ and $A_2$ as shorthand to denote events
$A_1:=\left\{T_1(t)<\frac{\log t}{(1+2\delta)K_{C'}}\right\}$ and
$A_2:=\left\{\lambda(T_1(t))\leq \frac{(1+\delta)\log t}{(1+2\delta)}\right\}$,
and by (\ref{eq:thm5-1}), we have
\begin{eqnarray}
\limsup_{t\rightarrow\infty}{\mathsf P}_{C_0}\left(A_1\cap A_2\right)>0.
\label{eq:thm5-2}
\end{eqnarray}

%Using $A_1$ and $A_2$ as shorthand to denote events
%$A_1:=\left\{T_1(t)<\frac{\log t}{(1+2\delta)K_{C'}}\right\}$ and
%$A_2:=\left\{\lambda(T_1(t))\leq \frac{(1+\delta)\log t}{(1+2\delta)}\right\}$,
\noindent The quantity $\EE_{C'}\left\{T_{inf}(t)\right\}$ can be rewritten as
follows. {\footnotesize
\begin{eqnarray}
&& {\mathsf E}_{C'}\left\{T_{inf}(t)\right\}\nonumber\\
&&\stackrel{(a)}{=}~ {\mathsf
 E}_{C'}\left\{T_{2}(t)\right\}\nonumber\\
&&\stackrel{(b)}{=}~  {\mathsf
 E}_{C'}\left\{t-T_{1}(t)\right\}\nonumber\\
&&\stackrel{(c)}{\geq}~ \left(t-\frac{\log t}{(1+2\delta)K_{C'}}\right){\mathsf
P}_{C'}\left(A_1
\right)\nonumber\\
&&\stackrel{(d)}{\geq}~\left(t-\frac{\log t}{(1+2\delta)K_{C'}}\right){\mathsf
P}_{C'}\left(A_1\cap A_2 \right) \nonumber\\
&&\stackrel{(e)}{\geq}~ \left(t-\frac{\log
t}{(1+2\delta)K_{C'}}\right)e^{-\frac{(1+\delta)\log
t }{1+2\delta}}{\mathsf P}_{C_0}\left(A_1\cap A_2 \right) \nonumber\\
&&\stackrel{(f)}{=}~ {\mathcal
O}\left(t^\frac{\delta}{1+2\delta}\right).\label{eq:thm5-3}
\end{eqnarray}}\cmpt
The equality marked $(a)$ follows from $M_{C'}=1$ and $(b)$ follows from the
fact that $T_1(t)+T_2(t)=t$. $(c)$ and $(d)$ follow from elementary probability
inequalities. $(e)$ follows from the change-of-measure formula and the
definition of $A_2$ in which $\lambda(T_1(t))\leq \frac{(1+\delta)\log
t}{(1+2\delta)}$. $(f)$ follows from simple arithmetic and
Eq.~(\ref{eq:thm5-2}).

The inequality (\ref{eq:thm5-3}) contradicts the assumption that
$\{\phi_\tau\}$ is uniformly good for both $C_0=(\theta_1,\theta_2)$ and
$C'=(\theta,\theta_2)$, and thus we have
\begin{eqnarray}
\lim_{t\rightarrow\infty} {\mathsf
P}_{C_0}\left(T_1(t)\geq\frac{(1-\epsilon)\log t}{K_{C'}}\right)=0, ~~\forall
\epsilon>0.\nonumber
\end{eqnarray}
By choosing the $\theta$ in $C'=(\theta,\theta_2)$ with the minimizing
configuration $\inf_{\theta>\theta_2}\sup_xI(\theta_1,\theta|x)$, we complete
the proof of the first statement of \Theorem{\ref{thm: lbd-X-depends-on-x}}.
The second statement in \Theorem{\ref{thm: lbd-X-depends-on-x}} can be obtained
by simply applying Markov's inequality and the first statement.
\end{proof}

%%%%%%%%%%%%%%%%%%%%%%%%%%%%%%%%%%%%%%%%%%%%%%%%%%%%%%%%%%%%%%%%%%%%%%%%%%%%%%%%%%%%%%%%%%
%
%   Analysis of the scheme of wait until good
%
%%%%%%%%%%%%%%%%%%%%%%%%%%%%%%%%%%%%%%%%%%%%%%%%%%%%%%%%%%%%%%%%%%%%%%%%%%%%%%%%%%%%%%%%%%
\section{Proof of \Theorem{\ref{thm: scheme-X-depends-on-x}} \label{app: pf-scheme-X-depends-on-x}}

We prove \Theorem{\ref{thm: scheme-X-depends-on-x}} by decomposing the inferior
sampling time instants into disjoint subsequences, each of which will be
discussed in separate lemmas respectively. For simplicity, throughout this
proof, we use $1\{\cd{1}(t)\}$ as shorthand for $1\{\mbox{\cd{1} is satisfied
at time $t$}\}$\footnote{``At time $t$" means after observing $X_t$ but before
the final decision $\phi_t$ is made. It is basically the moment when we are
performing the $\phi_t$-deciding algorithm.}, and use $\delta\nbd(G)$ to denote
the $\delta$-neighborhood of the distribution $G(x)$ on the $L^{\infty}$ space
of distributions.

Suppose $M_{C_0}=2$. To prove that for the $\{\phi_\tau\}$ in
\Algorithm{\ref{alg:X-depends-on-x}},
$\limsup_{t\rightarrow\infty}\frac{{\mathsf E}_{C_0}T_1(t)}{\log t}\leq
\frac{1}{\inf_{\theta>\theta_2}\max_xI(\theta_1,\theta|x)}$, we first note the
following:{\scriptsize
\begin{eqnarray}
&&T_1(t)\label{eq:all-terms-X-depends-on-x}\\
&&=~\sum_{\tau=1}^t1\{\phi_\tau=1\}\nonumber\\
&&=~\sum_{\tau=1}^t1\{\phi_\tau=1, \cd{0}(\tau)\}\nonumber\\
&&~~~~+\sum_{\tau=1}^t1\{\phi_\tau=1, \cd{1}(\tau)\}\nonumber\\
&&~~~~+\sum_{\tau=1}^t1\{\phi_\tau=1, \cd{2}(\tau)\}\nonumber\\
&&~~~~+\sum_{\tau=1}^t1\{\phi_\tau=1, \hat{C}_{\tau-1}=(\theta^\alpha,\theta^\beta), \theta^\alpha<\theta^\beta\neq\theta_2,\cd{3}(\tau)\}\nonumber\\
&&~~~~+\sum_{\tau=1}^t1\{\phi_\tau=1, \hat{C}_{\tau-1}=(\theta^\alpha,\theta^\beta), \theta_1\neq \theta^\alpha>\theta^\beta,\cd{3}(\tau)\}\nonumber\\
&&~~~~+\sum_{\tau=1}^t1\{\phi_\tau=1, \hat{C}_{\tau-1}=(\theta^\alpha,\theta^\beta), \theta_1\neq \theta^\alpha<\theta^\beta=\theta_2,\cd{3}(\tau)\}\nonumber\\
&&~~~~+\sum_{\tau=1}^t1\{\phi_\tau=1, \hat{C}_{\tau-1}=(\theta^\alpha,\theta^\beta), \theta_1=\theta^\alpha>\theta^\beta,\cd{3}(\tau)\}\nonumber\\
&&~~~~+\sum_{\tau=1}^t1\{\phi_\tau=1,
\hat{C}_{\tau-1}=(\theta^\alpha,\theta^\beta)=C_0=(\theta_1,\theta_2),\cd{3}(\tau)\}.\nonumber
\end{eqnarray}}\cmpt
These eight terms of the right-hand side of (\ref{eq:all-terms-X-depends-on-x})
will be treated separately in \Lemmas{\ref{lem:cond0} {\rm
through}~\ref{lem:term3e}}.

%%%%%%%%%%%%%%%%%%%%%%%%%%%%%%%%%%%%%%%%%%%%%%%%%%%%%%%%%%%%%%%%%%%%%%%%%%%%%%%%%%%%%%%%%%
%
%   Wait until good: proof of T0
%
%%%%%%%%%%%%%%%%%%%%%%%%%%%%%%%%%%%%%%%%%%%%%%%%%%%%%%%%%%%%%%%%%%%%%%%%%%%%%%%%%%%%%%%%%%
\begin{lemma}\label{lem:cond0}Suppose $M_{C_0}=2$, i.e., $\theta_1<\theta_2$.\footnote{There is no need to consider the case
$\theta_1=\theta_2$, since in that case, all allocation rules are optimal.}
Then
\begin{eqnarray}
\forall C_0\in{\mathbf{\Theta}}^2,&&\lim_{t\rightarrow\infty}{\mathsf
E}_{C_0}\left\{\sum_{\tau=1}^t1\{\phi_\tau=1,
\cd{0}(\tau)\}\right\}\nonumber\\
&&\leq~ \lim_{t\rightarrow\infty}{\mathsf
E}_{C_0}\left\{\sum_{\tau=1}^t1\{\cd{0}(\tau)\}\right\}<\infty.\nonumber
\end{eqnarray}
\end{lemma}
\begin{proof}
Let $T0:=\sum_{\tau=1}^\infty1\{\cd{0}(\tau)\}$. By the monotone convergence
theorem, it is equivalent to prove that ${\mathsf
E}_{C_0}\left\{T0\right\}<\infty$ for all $C_0$. By the definition of \cd{0},
we have{\footnotesize
\begin{eqnarray}
        &&{\mathsf P}_{C_0}(T0=t)\nonumber\\
        &&\leq~\sum_{x\in{\mathbf X}}{\mathsf P}_{G}(X_t=x, \forall \tau<t\mbox{~and~}\tau\equiv \mdt{t}, X_\tau\neq x)\nonumber\\
        &&=~\sum_{x\in{\mathbf X}}{\mathsf P}_{G}(X_t=x)(1-{\mathsf P}_{G}(X=x))^{[\frac{t-1}{2}]}\nonumber\\
        &&\leq~\left(1-\min_x {\mathsf P}_{G}(X_t=x)\right)^{[\frac{t-1}{2}]}.\nonumber
\end{eqnarray}}\cmpt
By directly computing the expectation, we obtain ${\mathsf
E}_{C_0}\{T0\}<\infty$.
\end{proof}

%%%%%%%%%%%%%%%%%%%%%%%%%%%%%%%%%%%%%%%%%%%%%%%%%%%%%%%%%%%%%%%%%%%%%%%%%%%%%%%%%%%%%%%%%%
%
%   Wait until good: proof of term 1
%
%%%%%%%%%%%%%%%%%%%%%%%%%%%%%%%%%%%%%%%%%%%%%%%%%%%%%%%%%%%%%%%%%%%%%%%%%%%%%%%%%%%%%%%%%%
\begin{lemma} \label{lem:cond1}Suppose $M_{C_0}=2$, i.e., $\theta_1<\theta_2$.
Then
\begin{eqnarray}
&&\lim_{t\rightarrow\infty}{\mathsf
E}_{C_0}\left\{\sum_{\tau=1}^t1\{\phi_\tau=1,
\cd{1}(\tau)\}\right\}\nonumber\\
&&\leq~ \lim_{t\rightarrow\infty}{\mathsf
E}_{C_0}\left\{\sum_{\tau=1}^t1\{\cd{1}(\tau)\}\right\}<\infty.\nonumber
\end{eqnarray}
\end{lemma}

\begin{proof}
We define $L_X(t|\cd{1})$ as the empirical distribution of $X_\tau$ at those
time instants $\tau\leq t$ for which \cd{1} is satisfied. We then have
{\footnotesize
\begin{eqnarray}
&&\sum_{\tau=1}^t1\{\cd{1}(\tau)\}\nonumber\\
        &&=~\sum_{\tau=1}^t1\{\cd{1}(\tau),L_X(\tau|\cd{1})\in\delta\nbd(G)\}\nonumber\\
                &&~~~~~+\sum_{\tau=1}^t1\{\cd{1}(\tau),L_X(\tau|\cd{1})\notin\delta\nbd(G)\}.
                \label{eq:pf-cond1}
\end{eqnarray}}\cmpt
By Sanov's theorem on finite alphabets (see \cite{DemboZeitouni90}), each term
in the second sum is exponentially upper bounded w.r.t.\ $\tau$, which implies
the bounded expectation of the second sum. For the first sum, we have
{\footnotesize
\begin{eqnarray}
&&\sum_{\tau=1}^t1\{\cd{1}(\tau),L_X(\tau|\cd{1})\in\delta\nbd(G)\}\nonumber\\
        &&\leq~\sum_{\tau=1}^\infty1
            \{\exists i,x,\st L_i^x(\tau-1)\notin\epsilon\nbd(F_{\theta_i}(\cdot|x)), \nonumber\\
            &&~~~~~~~~~~~~~~~~~~~\mbox{and }L_X(\tau|\cd{1})\in \delta\nbd(G)\}\nonumber\\
        &&\leq~\sum_{x}\sum_{i=1}^2\sum_{\tau=1}^\infty
            1\{ L_i^x(\tau-1)\notin\epsilon\nbd(F_{\theta_i}(\cdot|x)), \nonumber\\
            &&~~~~~~~~~~~~~~~~~~~~~~L_X(\tau|\cd{1})\in
            \delta\nbd(G)\}\label{eq:change-time1}\\
        &&\leq~\sum_{x}\sum_{i=1}^2
        \sum_{\tau'=1}^{\infty}
                1\{\exists n\geq\left[\frac{\tau'{\mathsf P}_{G}(X=x)(1-\delta)-1}{2}\right],\nonumber\\
                &&~~~~~~~~~~~~~~~~~~~~~~~~\st \rho(L_i^x(n),F_{\theta_i}(\cdot|x))>\epsilon\}.\label{eq:change-time2}
\end{eqnarray}}\cmpt
The first inequality comes from extending the finite sum to the infinite sum
and the definition of \cd{1}. The second inequality comes from the union bound.
 The third inequality comes from the following three
steps. First we change the summation index from the time variable $\tau$ to
$\tau'$, which specifies that it is the $\tau'$-th time that the condition in
(\ref{eq:change-time1}) is satisfied. (Note: by definition, $\tau\geq \tau'$.)
Second, by $L_X(\tau|\cd{1})\in
            \delta\nbd(G)$, there must be at least $\tau'{\mathsf P}_{G}(X=x)(1-\delta)$ time instants that
            $X_s=x$, $s\leq \tau$, which guarantees  we have enough access to the bandit machine $x$. And finally, by the definition
of \cd{1} in \Algorithm{\ref{alg:X-depends-on-x}}, at the $\tau'$-th time of
satisfaction, the sample size $n$ must be greater than
$\left[\frac{\tau'{\mathsf P}_{C_0}(X=x)(1-\delta)-1}{2}\right]$. By slightly
abusing the notation $L_i^x(t)$ with $L_i^x(n)$, where $n$ represents the
sample size $T^x_i(t)$ rather than the current time $t$, we obtain the third
inequality.

Remark: this change-of-index transformation will be used extensively throughout
the proofs in this section.

By Sanov's theorem on $\mathbb R$ (\Theorem{\ref{thm:Sanov}}), the probability
of each term in (\ref{eq:change-time2}) is exponentially upper bounded w.r.t.\
$\tau'$, which implies that the summation has bounded expectation. By
(\ref{eq:pf-cond1}), the proof of \Lemma{\ref{lem:cond1}} is then complete.
\end{proof}

%%%%%%%%%%%%%%%%%%%%%%%%%%%%%%%%%%%%%%%%%%%%%%%%%%%%%%%%%%%%%%%%%%%%%%%%%%%%%%%%%%%%%%%%%%
%
%   Wait until good: proof of term 2
%
%%%%%%%%%%%%%%%%%%%%%%%%%%%%%%%%%%%%%%%%%%%%%%%%%%%%%%%%%%%%%%%%%%%%%%%%%%%%%%%%%%%%%%%%%%
\begin{lemma}
\label{lem:cond2}Suppose $M_{C_0}=2$, i.e., $\theta_1<\theta_2$. Then
\begin{eqnarray}
\lim_{t\rightarrow\infty}{\mathsf E}_{C_0}\left\{\sum_{\tau=1}^t1\{\phi_\tau=1,
\cd{2}(\tau)\}\right\}<\infty.\nonumber
\end{eqnarray}
\end{lemma}
\begin{proof}
By the assumption $\theta_1<\theta_2$, we have{\footnotesize
\begin{eqnarray}
&&\sum_{\tau=1}^t1\{\phi_\tau=1,
            \cd{2}(\tau)\}\nonumber\\
        &&=~\sum_{\tau=1}^t1\{\hat{C}_{\tau-1}=(\theta^\alpha,\theta^\beta), \theta^\alpha=\bar{\theta}\}\nonumber\\
        &&=~\sum_{\tau=1}^t1\{\hat{C}_{\tau-1}=(\theta^\alpha,\theta^\beta),\theta^\alpha=\bar{\theta}, L_X(\tau|\theta^\alpha=\bar{\theta})\in\delta\nbd(G)\}\nonumber\\
                &&~~~~+\sum_{\tau=1}^t1\{\hat{C}_{\tau-1}=(\theta^\alpha,\theta^\beta),\theta^\alpha=\bar{\theta},L_X(\tau|\theta^\alpha=\bar{\theta})\notin\delta\nbd(G)\}\nonumber
\end{eqnarray}}\cmpt
By Sanov's theorem on finite alphabets, each term in the second sum is
exponentially upper bounded w.r.t.\ $\tau$, which implies the bounded
expectation of the second sum. For the first sum, we have {\footnotesize
\begin{eqnarray}
&&\sum_{\tau=1}^t1\{\hat{C}_{\tau-1}=(\theta^\alpha,\theta^\beta),\theta^\alpha=\bar{\theta}, L_X(\tau|\theta^\alpha=\bar{\theta})\in\delta\nbd(G)\}\nonumber\\
        &&\leq~\sum_{\tau=1}^\infty1\{\hat{C}_{\tau-1}=(\theta^\alpha,\theta^\beta),\theta^\alpha=\bar{\theta}, \exists x,\st \nonumber\\
        &&~~~~~~~~~~~~~~\rho(L_1^x(\tau-1), F_{\theta_1}(\cdot|x))>\epsilon,L_X(\tau|\theta^\alpha=\bar{\theta})\in\delta\nbd(G)\}\nonumber\\
        &&\leq~\sum_{x}\sum_{\tau'=1}^{\infty}1\{\exists n\geq\left[ \tau'{\mathsf P}_{G}(X=x)(1-\delta)-1\right],\st\nonumber\\
        &&~~~~~~~~~~~~~~~~~~~~~~~~~~~~~~~~~~\rho(L_1^x(n),F_{\theta_1}(\cdot|x))>\epsilon\}.\nonumber
\end{eqnarray}}

 By extending the finite sum to the infinite sum, we
obtain the first inequality. By the definition of \cd{2} in
\Algorithm{\ref{alg:X-depends-on-x}} and using exactly the same reasoning used
in going from (\ref{eq:change-time1}) to (\ref{eq:change-time2}), we obtain the
second inequality. By Sanov's theorem, each term in the above sum is
exponentially upper bounded w.r.t.\ $\tau'$. Thus it follows that the
expectation of the first sum is also finite, which completes the proof.
\end{proof}

%%%%%%%%%%%%%%%%%%%%%%%%%%%%%%%%%%%%%%%%%%%%%%%%%%%%%%%%%%%%%%%%%%%%%%%%%%%%%%%%%%%%%%%%%%
%
%   Wait until good: proof of term 3a
%
%%%%%%%%%%%%%%%%%%%%%%%%%%%%%%%%%%%%%%%%%%%%%%%%%%%%%%%%%%%%%%%%%%%%%%%%%%%%%%%%%%%%%%%%%%
\begin{lemma}
\label{lem:term3a}Suppose $M_{C_0}=2$, i.e., $\theta_1<\theta_2$. Then
{\footnotesize
\begin{eqnarray}
&&\lim_{t\rightarrow\infty}{\mathsf
E}_{C_0}\left\{\sum_{\tau=1}^t1\left\{\phi_\tau=1,
\hat{C}_{\tau-1}=(\theta^\alpha,\theta^\beta),\right.\right.\nonumber\\
&&~~~~~~~~~~~~~~~~~~~\left.\left.\theta^\alpha<\theta^\beta\neq\theta_2,
\cd{3}(\tau)\right\}\right\}\nonumber\\
&&\leq~\lim_{t\rightarrow\infty}{\mathsf E}_{C_0}\left\{\sum_{\tau=1}^t1\left\{
\hat{C}_{\tau-1}=(\theta^\alpha,\theta^\beta),\right.\right.\nonumber\\
&&~~~~~~~~~~~~~~~~~~~\left.\left.\theta^\alpha<\theta^\beta\neq\theta_2,
\cd{3}(\tau)\right\}\right\}\nonumber\\
&&<~\infty.\nonumber
\end{eqnarray}}
\end{lemma}

\begin{proof}
We have {\scriptsize
\begin{eqnarray}
&&\sum_{\tau=1}^t1\left\{
\hat{C}_{\tau-1}=(\theta^\alpha,\theta^\beta),\theta^\alpha<\theta^\beta\neq\theta_2, \cd{3}(\tau)\right\}\nonumber\\
        &&=~\sum_{(\theta,\vartheta):\theta<\vartheta\neq\theta_2}\sum_{\tau=1}^t 1\{\hat{C}_{\tau-1}=(\theta,\vartheta), \cd{3}(\tau)\}\nonumber\\
        &&\leq~2\sum_{(\theta,\vartheta):\theta<\vartheta\neq\theta_2}\sum_{\tau=1}^t 1\{\hat{C}_{\tau-1}=(\theta,\vartheta), \cd{3a}(\tau)\}\nonumber\\
        &&=~2\sum_{(\theta,\vartheta):\theta<\vartheta\neq\theta_2}\sum_{x}\sum_{\tau=1}^t
                1\{X_\tau=x,\hat{C}_{\tau-1}=(\theta,\vartheta), \cd{3a}(\tau)\}\nonumber\\
        &&\leq~2\sum_{(\theta,\vartheta):\theta<\vartheta\neq\theta_2}\sum_x \sum_{\tau=1}^\infty
                1\{ \rho(L_2^x(\tau-1),F_{\theta_2}(\cdot|x))>\epsilon, \cd{3a}(\tau)\}\nonumber\\
        &&\leq~2\sum_{(\theta,\vartheta):\theta<\vartheta\neq\theta_2}\sum_x\sum_{\tau'=1}^\infty
            1\{\exists n\geq \left[\tau'-1\right],\st\nonumber\\
            &&~~~~~~~~~~~~~~~~~~~~~~~~~~~~~~~~~~~~~~~~~~~~~~~~~~~~ \rho(L_2^x(n),F_{\theta_2}(\cdot|x))>\epsilon\}.\nonumber
\end{eqnarray}}

The first equality follows from conditioning on the event that the exact value
of the estimate $\hat{C}_{\tau-1}$ is some configuration pair
$(\theta,\vartheta)$. The first inequality follows from the definition of
\cd{3a} in \Algorithm{\ref{alg:X-depends-on-x}}, where double the number of
time instants with odd \ctr{($\hat{C}$)} will be larger than the total number
of times that \cd{3} is satisfied. The second equality follows from
conditioning on the value of $X_\tau$.  The second inequality follows from the
condition that the second coordinate of the estimate,  $\vartheta\neq\theta_2$,
and then extending the finite sum to the infinite sum. The third inequality
follows from the definition of \cd{3a} and changing the time index to $\tau'$,
similarly to the reasoning in (\ref{eq:change-time1})--(\ref{eq:change-time2}).
By Sanov's theorem, each term is exponentially upper bounded w.r.t.\ $\tau'$,
and thus the entire sum has bounded expectation. The proof is thus complete.
\end{proof}

\begin{corollary}\label{lem:term3b}
By the symmetry of $\{\phi_\tau\}$, we have {\scriptsize
\begin{eqnarray}
&&\lim_{t\rightarrow\infty}\nonumber\\
&&{\mathsf E}_{C_0}\left\{ \sum_{\tau=1}^t1\left\{\phi_\tau=1,
\hat{C}_{\tau-1}=(\theta^\alpha,\theta^\beta),\theta_1\neq\theta^\alpha>\theta^\beta,
\cd{3}(\tau)\right\}\right\}\nonumber\\
&&<~\infty.\nonumber
\end{eqnarray}}
\end{corollary}

%%%%%%%%%%%%%%%%%%%%%%%%%%%%%%%%%%%%%%%%%%%%%%%%%%%%%%%%%%%%%%%%%%%%%%%%%%%%%%%%%%%%%%%%%%
%
%   Wait until good: proof of term 3c
%
%%%%%%%%%%%%%%%%%%%%%%%%%%%%%%%%%%%%%%%%%%%%%%%%%%%%%%%%%%%%%%%%%%%%%%%%%%%%%%%%%%%%%%%%%%
\begin{lemma}
\label{lem:term3c}Suppose $M_{C_0}=2$, i.e., $\theta_1<\theta_2$. Then
\begin{eqnarray}
&&\lim_{t\rightarrow\infty}{\mathsf
E}_{C_0}\left\{\sum_{\tau=1}^t1\left\{\phi_\tau=1,
\hat{C}_{\tau-1}=(\theta^\alpha,\theta^\beta)\right.\right.,\nonumber\\
&&~~~~~~~~~~~~~\left.\left.\theta_1\neq\theta^\alpha<\theta^\beta=\theta_2,
\cd{3}(\tau)\right\}\right\}<\infty.\nonumber
\end{eqnarray}
\end{lemma}
\begin{proof} We have{\footnotesize
\begin{eqnarray}
&&\sum_{\tau=1}^t1\left\{\phi_\tau=1,\hat{C}_{\tau-1}=(\theta^\alpha,\theta^\beta),\theta_1\neq\theta^\alpha<\theta^\beta=\theta_2,
\cd{3}(\tau)\right\}\nonumber\\
        &&=~\sum_{\theta:\theta_1\neq\theta<\theta_2}\sum_{\tau=1}^t
            1\{\phi_\tau=1, \hat{C}_{\tau-1}=(\theta,\theta_2),\cd{3}(\tau)\}\nonumber\\%\label{eq: diff on term3c}\\
        &&=~\sum_{\theta:\theta_1\neq\theta<\theta_2}\sum_{\tau=1}^t
                (1\{\phi_\tau=1, \hat{C}_{\tau-1}=(\theta,\theta_2),\cd{3b1a1}(\tau)\}\nonumber\\
                &&~~~~~~~~~~~~~~~~~+1\{\phi_\tau=1, \hat{C}_{\tau-1}=(\theta,\theta_2),\cd{3b1a2}(\tau)\})\nonumber\\
        &&\leq~\sum_{\theta_1\neq\theta<\vartheta=\theta_2}\sum_{\theta':\theta'>\theta_2}\nonumber\\
                &&~~~~(\sum_{\tau=1}^\infty1\{\phi_\tau=1, \hat{C}_{\tau-1}=(\theta,\theta_2), \theta^*=\theta', \cd{3b1a1}(\tau)\}\nonumber\\
                &&~~~~+\sum_{\tau=1}^\infty1\{\phi_\tau=1, \hat{C}_{\tau-1}=(\theta,\theta_2),\theta^*=\theta', \cd{3b1a2}(\tau)\})\label{eq:tm3c-1}\\
        &&\leq~\sum_{\theta:\theta_1\neq\theta<\theta_2}\sum_{\theta':\theta'>\theta_2}\sum_{\tau'=1}^\infty\nonumber\\
                &&~~~~(4(\tau'+1)^2-(2\tau')^2)\cdot
                1\{\exists n\geq [\tau'-1], \st\nonumber\\
        &&~~~~~~~~~~~~~~~~~~~~~~~~~~~~~~~~\rho(L_1^{x^*(\theta')}(n),F_{\theta_1}(\cdot|x^*(\theta')))>\epsilon\}.\label{eq:tm3c-2}
\end{eqnarray}}

The second equality comes from the fact that the scheme samples the inferior
arm only when either \cd{3b1a1} or \cd{3b1a2} is satisfied. For the first
inequality, we condition on $\theta^*$ and extend to the infinite sum. For the
last inequality, we change the time index to $\tau'$, which specifies the
$\tau'$-th satisfaction of \cd{3b1a1}, so that we can upper bound the first sum
of (\ref{eq:tm3c-1}). The reason we have a multiplication factor
$(4(\tau'+1)^2-(2\tau')^2)$ in front of the indicator function is in order to
upper bound the second sum of (\ref{eq:tm3c-1}), concerning \cd{3b1a2},
simultaneously.

To obtain this result, we note that between the consecutive times $\tau'$ and
$\tau'+1$, at which \cd{3b1a1} is satisfied and arm~1 is pulled, the number of
times that \cd{3b1a2} is satisfied and arm~1 is pulled cannot exceed
$(2(\tau'+1))^2-(2\tau')^2-1$, which is because of the algorithm involving
\ctr{($\hat{C}_t,\theta^*$)} in Line~\readalgline{cond3b1a1}. Multiplying the
factor $(4(\tau'+1)^2-(2\tau')^2)$, we simultaneously bound these two sums.

By Sanov's theorem, the expectation of the indicator in (\ref{eq:tm3c-2}) is
exponentially upper bounded w.r.t.\ $\tau'$. As a result, the entire sum will
have bounded expectation, which in turn completes the proof.
\end{proof}

%%%%%%%%%%%%%%%%%%%%%%%%%%%%%%%%%%%%%%%%%%%%%%%%%%%%%%%%%%%%%%%%%%%%%%%%%%%%%%%%%%%%%%%%%%
%
%   Wait until good: proof of term 3d
%
%%%%%%%%%%%%%%%%%%%%%%%%%%%%%%%%%%%%%%%%%%%%%%%%%%%%%%%%%%%%%%%%%%%%%%%%%%%%%%%%%%%%%%%%%%
\begin{lemma}
\label{lem:term3d}Suppose $M_{C_0}=2$, i.e., $\theta_1<\theta_2$. Then
\begin{eqnarray}
\lim_{t\rightarrow\infty}{\mathsf
E}_{C_0}\left\{\sum_{\tau=1}^t1\left\{\phi_\tau=1,\hat{C}_{\tau-1}=(\theta^\alpha,\theta^\beta),\right.\right.\nonumber\\
\left.\left.\theta_1=\theta^\alpha>\theta^\beta,
\cd{3}(\tau)\right\}\right\}<\infty.\nonumber
\end{eqnarray}
\end{lemma}
\begin{proof}
We have, {\footnotesize
\begin{eqnarray}
&&\sum_{\tau=1}^t1\left\{\phi_\tau=1,\hat{C}_{\tau-1}=(\theta^\alpha,\theta^\beta),\theta_1=\theta^\alpha>\theta^\beta, \cd{3}(\tau)\right\}\nonumber\\
        &&=~\sum_{\vartheta:\theta_1>\vartheta}\sum_{\tau=1}^t 1\{\phi_\tau=1, \hat{C}_{\tau-1}=(\theta_1,\vartheta),\cd{3}(\tau)\}\nonumber\\
        &&\leq~\sum_{\vartheta:\theta_1>\vartheta}\sum_{\tau=1}^t 1\{\hat{C}_{\tau-1}=(\theta_1,\vartheta),\cd{3}(\tau)\}\nonumber\\
        &&\leq~\sum_{\vartheta:\theta_1>\vartheta}\left(2\sum_{\tau=1}^t1\{\hat{C}_{\tau-1}=(\theta_1,\vartheta),\cd{3b}(\tau)\}+1\right)\nonumber\\
        &&=~\#\{\vartheta\in{\mathbf \Theta}:\vartheta<\theta_1\}+\sum_{\vartheta:\theta_1>\vartheta}\nonumber\\
        &&~~~~\left(2\sum_{\theta':\theta'>\theta_1}\sum_{\tau=1}^t
            1\{\hat{C}_{\tau-1}=(\theta_1,\vartheta),\theta^*=\theta',\cd{3b}(\tau)\}\right)\nonumber\\
        &&=~\#\{\vartheta\in{\mathbf \Theta}:\vartheta<\theta_1\}\nonumber\\
                &&~~~~+2\sum_{\vartheta:\theta_1>\vartheta}\sum_{\theta':\theta'>\theta_1}\sum_{\tau=1}^t\nonumber\\
                &&~~~~~~~~~~(1\{\hat{C}_{\tau-1}=(\theta_1,\vartheta),\theta^*=\theta',\nonumber\\
                &&~~~~~~~~~~~~~~~~~~~\cd{3b}(\tau), L_X(\tau|\cd{3b})\in\delta\nbd(G)\}\nonumber\\
                &&~~~~~~~~~~~~+1\{\hat{C}_{\tau-1}=(\theta_1,\vartheta),\theta^*=\theta',\nonumber\\
                &&~~~~~~~~~~~~~~~~~~~\cd{3b}(\tau), L_X(\tau|\cd{3b})\notin\delta\nbd(G)\}).\label{eq:tm3d-1}
\end{eqnarray}}

The first inequality follows from Line~\readalgline{cond3b} in
\Algorithm{\ref{alg:X-depends-on-x}}, where \cd{3b} is satisfied once after two
times of \cd{3} satisfaction. The last two equalities come from conditioning on
$\theta^*$ and $L_X(\tau|\cd{3b})$. By Sanov's theorem on finite alphabets, the
terms of the second sum in (\ref{eq:tm3d-1}) are exponentially upper bounded
and the entire sum thus has bounded expectation. For the first sum, we have
\begin{eqnarray}
&&\sum_{\tau=1}^t1\{\hat{C}_{\tau-1}=(\theta_1,\vartheta),\theta^*=\theta',\nonumber\\
&&~~~~~~~~~~~~~\cd{3b}(\tau), L_X(\tau|\cd{3b})\in\delta\nbd(G)\}\nonumber\\
        &&\leq~\frac{1}{{\mathsf P}_{G}(X=x^*(\theta_1,\vartheta))(1-\delta)}\cdot\nonumber\\
        &&~~~~~\sum_{\tau=1}^t
        1\{\hat{C}_{\tau-1}=(\theta_1,\vartheta),\theta^*=\theta',\cd{3b1}(\tau)\}.\nonumber\\
        &&\label{eq:tm3d-3}
\end{eqnarray}
This inequality follows from the fact that once $L_X$ falls into the
$\delta$-nbd($G$), the total number of time instants can be upper bounded by
the number of instants when $X_\tau=x^*(\theta_1,\vartheta)$, over ${\mathsf
P}_G(X=x^*(\theta_1,\vartheta))(1-\delta)$. To show {\footnotesize
\begin{eqnarray}
\lim_{t\rightarrow\infty}{\mathsf E
}_{C_0}\left\{\sum_{\tau=1}^t1\{\hat{C}_{\tau-1}=(\theta_1,\vartheta),\theta^*=\theta',\cd{3b1}(\tau)\}\right\}<\infty,\nonumber
\end{eqnarray}}\cmpt
we further decompose the expectand into {\footnotesize
\begin{eqnarray}
&&\sum_{\tau=1}^t1\{\hat{C}_{\tau-1}=(\theta_1,\vartheta),\theta^*=\theta',\cd{3b1}(\tau)\}\nonumber\\
&&=~\sum_{\tau=1}^t1\{\hat{C}_{\tau-1}=(\theta_1,\vartheta),\theta^*=\theta',\cd{3b1a}(\tau)\}\nonumber\\
&&~~~~+\sum_{\tau=1}^t1\{\hat{C}_{\tau-1}=(\theta_1,\vartheta),\theta^*=\theta',\cd{3b1b}(\tau)\}.\label{eq:tm3d-2}
\end{eqnarray}}

For the first sum in (\ref{eq:tm3d-2}), under the assumption
$\theta_1>\vartheta$, we can write {\footnotesize
\begin{eqnarray}
&&\sum_{\tau=1}^t1\{\hat{C}_{\tau-1}=(\theta_1,\vartheta),\theta^*=\theta',\cd{3b1a}(\tau)\}\nonumber\\
&&\leq~\sum_{\tau=1}^\infty1\{\hat{C}_{\tau-1}=(\theta_1,\vartheta),\theta^*=\theta',\cd{3b1a1}(\tau)\}\nonumber\\
&&~~~~+\sum_{\tau=1}^\infty1\{\hat{C}_{\tau-1}=(\theta_1,\vartheta),\theta^*=\theta',\cd{3b1a2}(\tau)\}\nonumber\\
&&\leq~1+2\sum_{\tau=1}^\infty1\{\hat{C}_{\tau-1}=(\theta_1,\vartheta),\theta^*=\theta',\cd{3b1a2}(\tau)\}\nonumber\\
&&\leq~1+2\sum_{\tau'=1}^\infty1\{\rho(\exists n\geq[\tau-1], \st\nonumber\\
&& ~~~~~~~~~~~~~~~~~~
\rho(L_2^{x^*(\theta_1,\vartheta)}(n),F_{\theta_2}(\cdot|x^*(\theta_1,\vartheta)))>\epsilon\}.\label{eq:tm3d-4}
\end{eqnarray}}

The first inequality comes from conditioning on the sub-conditions \cd{3b1a1}
and \cd{3b1a2}, and extending to the infinite sums. Let ${\mathsf{SQ}}_n$
denote the set of perfectly squared integers in $\{1,\cdots, n\}$. The second
inequality is from the definition of \cd{3b1a1} in
\Algorithm{\ref{alg:X-depends-on-x}} and the fact that $\forall n\in
{\mathbb{N}}$, $|{\mathsf{SQ}}_n|$ is no larger than
$1+|\{1,\cdots,n\}\backslash{\mathsf{SQ}}_n|$. The third inequality comes from
the fact that by definition, under \cd{3b1a2}, $\phi_\tau=2$, and changing the
time index to $\tau'$, the number of satisfaction times. By Sanov's theorem on
$\mathbb R$, the above has bounded expectation.

For the second sum of (\ref{eq:tm3d-2}), with the condition
$\theta_1>\vartheta$ {\scriptsize
\begin{eqnarray}
&&\sum_{\tau=1}^t1\{\hat{C}_{\tau-1}=(\theta_1,\vartheta),\theta^*=\theta',\cd{3b1b}(\tau)\}\nonumber\\
&&\leq~1+\sum_{\tau=1}^{t-1} 1\{\hat{C}_{\tau}=(\theta_1,\vartheta), \theta^*=\theta',\Lambda_{\tau}(\hat{C}_{\tau}, \theta')>\tau(\log\tau)^2\}\nonumber\\
&&\leq~1+\sum_{\tau=1}^{t-1} 1\{\hat{C}_{\tau}=(\theta_1,\vartheta), \Lambda_\tau(\hat{C}_\tau,\theta_2)>\tau(\log\tau)^2\}\nonumber\\
&&\leq~1+\sum_{\tau=1}^{t-1}
   1\{\prod_{m=1}^{T_2^{x^*(\theta_1,\vartheta)}(\tau)}\frac{F_{\vartheta}
   (dY_{\tau_{x^*}(m)}^{2}|x^*(\theta_1,\vartheta))}
   {F_{\theta_2}
   (dY_{\tau_{x^*}(m)}^{2}|x^*(\theta_1,\vartheta))}>\tau(\log\tau)^2,\}\nonumber\\
&&\leq~1+\sum_{\tau=1}^{t-1}
   1\{\exists n\leq \tau,\st\prod_{m=1}^{n}\frac{F_{\vartheta}
   (dY_{m|x^*}^{2}|x^*(\theta_1,\vartheta))}
   {F_{\theta_2}
   (dY_{m|x^*}^{2}|x^*(\theta_1,\vartheta))}>\tau(\log\tau)^2\},\nonumber
\end{eqnarray}}\cmpt
where $Y^2_{m|x^*}$ is the reward of arm~2 at the $m$-th time that
$X_s=x^*(\theta_1,\vartheta)$ and $\phi_s=2$.
 The first inequality follows from
focusing only on the $\Lambda_{\tau-1}(\hat{C}_{\tau-1},\theta')$ condition in
\cd{3b1b} and then shifting the time index $\tau$. The second inequality
follows by replacing the minimum achieving $\theta'$ with $\theta_2$. The third
inequality follows from expressing $\Lambda_\tau$ using its definition. The
fourth inequality follows from the set relationship, where $n$ is
$T_2^{x^*(\theta_1,\vartheta)}(\tau)$, the number of time instants that the
side information $X_s=x^*(\theta_1,\vartheta)$ and $\phi_s=2$, for $s\leq
\tau$.

We first note that
$\prod_{m=1}^n\frac{F_{\vartheta}(dY^2_{m|x^*}|x^*(\theta_1,\vartheta))}
{F_{\theta_2}(dY^2_{m|x^*}|x^*(\theta_1,\vartheta))}$ is a positive martingale
with expectation 1, when being considered  under distribution
$F_{\theta_2}(\cdot|x^*(\theta_1,\vartheta))$. By Doob's maximal inequality, we
have{\footnotesize
\begin{eqnarray}
{\mathsf P}_{C_0}\left(\exists n\leq \tau, \prod_{m=1}^n\frac{F_{\vartheta}
(dY_{m|x^*}^{2}|x^*(\theta_1,\vartheta))}{F_{\theta_2}
(dY_{m|x^*}^{2}|x^*(\theta_1,\vartheta))}>\tau(\log\tau)^2\right)
\leq\frac{1}{\tau(\log \tau)^2},\nonumber
\end{eqnarray}}\cmpt
and thus the expectation is bounded, i.e., {\footnotesize
\begin{eqnarray}
&&{\mathsf
E}_{C_0}\left\{\sum_{\tau=1}^t1\{\hat{C}_{\tau-1}=(\theta_1,\vartheta),\theta^*=\theta',\cd{3b1b}(\tau)\}\right\}\nonumber\\
&&\leq~1+\sum_{\tau=1}^\infty\frac{1}{\tau(\log\tau)^2}<\infty.
\label{eq:tm3d-5}
\end{eqnarray}}\cmpt
By (\ref{eq:tm3d-1}), (\ref{eq:tm3d-3}), (\ref{eq:tm3d-2}), (\ref{eq:tm3d-4}),
and (\ref{eq:tm3d-5}), \Lemma{\ref{lem:term3d}} is proved.
\end{proof}

%%%%%%%%%%%%%%%%%%%%%%%%%%%%%%%%%%%%%%%%%%%%%%%%%%%%%%%%%%%%%%%%%%%%%%%%%%%%%%%%%%%%%%%%%%
%
%   Wait until good: proof of term 3e
%
%%%%%%%%%%%%%%%%%%%%%%%%%%%%%%%%%%%%%%%%%%%%%%%%%%%%%%%%%%%%%%%%%%%%%%%%%%%%%%%%%%%%%%%%%%
\begin{lemma}
\label{lem:term3e}Suppose $M_{C_0}=2$, i.e., $\theta_1<\theta_2$. Then
{\footnotesize
\begin{eqnarray}
&&\limsup_{t\rightarrow\infty}\frac{1}{\log t}\cdot\nonumber\\
&&{\mathsf E}_{C_0}\left\{\sum_{\tau=1}^t1\left\{\phi_\tau=1,
\hat{C}_{\tau-1}=(\theta^\alpha,\theta^\beta)=(\theta_1,\theta_2),
\cd{3}(\tau)\right\}\right\}\nonumber\\
&&\leq~\frac{1}{\inf_{\theta>\theta_2}\max_xI(\theta_1,\theta|x)}.\nonumber
\end{eqnarray}}
\end{lemma}
\begin{proof}
By the definition of $\{\phi_\tau\}$, especially of \cd{3b1a}, we have
{\scriptsize
\begin{eqnarray}
&&{\mathsf
    E}_{C_0}\left\{\sum_{\tau=1}^t1\left\{\phi_\tau=1,
    \hat{C}_{\tau-1}=(\theta^\alpha,\theta^\beta)=C_0=(\theta_1,\theta_2),
    \cd{3}(\tau)\right\}\right\}\nonumber\\
&&\leq~{\mathsf
    E}_{C_0}\left\{\sum_{\tau=1}^t1\{\hat{C}_{\tau-1}=C_0, \cd{3b1a}(\tau)\}\right\}\nonumber\\
&&\leq~{\mathsf E}_{C_0}\left\{\sup\{1\leq n\leq t-1:\right.\nonumber\\
&&~~~~~~~~~~~~~~~~\left.
    \min_{\theta>\theta_2}\prod_{m=1}^n\frac{F_{\theta_1}
    (dY_{m|x^*}^{1}|x^*(C_0))}{F_{\theta}
    (dY_{m|x^*}^{1}|x^*(C_0))}\leq
    t(\log t)^2\}\right\}\nonumber\\
&&\leq~{\mathsf E}_{C_0}\left\{\sup\{1\leq n<\infty:\right.\nonumber\\
&&~~~~~~~~~~~~~~~~\left.
    \min_{\theta>\theta_2}\prod_{m=1}^n\frac{F_{\theta_1}
    (dY_{m|x^*}^{1}|x^*(C_0))}{F_{\theta}
    (dY_{m|x^*}^{1}|x^*(C_0))}\leq
    t(\log t)^2\}\right\}\nonumber\\
&&=~{\mathsf E}_{C_0}\left\{\max_{\theta>\theta_2}\sup\{1\leq
n<\infty:\right.\nonumber\\
&&~~~~~~~~~~~~~~~~\left.    \prod_{m=1}^n\frac{F_{\theta_1}
    (dY_{m|x^*}^{1}|x^*(C_0))}{F_{\theta}
    (dY_{m|x^*}^{1}|x^*(C_0))}\leq
    t(\log t)^2\}\right\}\nonumber\\
&&=~{\mathsf E}_{C_0}\left\{\max_{\theta>\theta_2}\sum_{s=1}^\infty
1\{\inf_{n\geq s}
    \prod_{m=1}^n\frac{F_{\theta_1}
    (dY_{m|x^*}^{1}|x^*(C_0))}{F_{\theta}
    (dY_{m|x^*}^{1}|x^*(C_0))}\leq
    t(\log t)^2\}\right\},\nonumber
\end{eqnarray}}\cmpt
where $Y^1_{m|x^*}$ denotes the reward of the $m$-th time that arm~1  of the
sub-bandit machine $X_\tau=x^*(C_0)$ is pulled.
 The first inequality follows because, by definition, only when \cd{3b1a} is
satisfied can $\phi_\tau=1$, given $\hat{C}_{\tau-1}=C_0$. The second
inequality is obtained by focusing on the sub-condition
$\Lambda_\tau(\hat{C}_\tau,\theta)$ in \cd{3b1a}, and letting
$n=T_1^{x^*}(t-1)$ be the number of time instants when arm~1 is pulled and
$X_\tau=x^*(C_0)$.
 The third inequality comes from
extending the upper bound of $n$ from $t-1$ to $\infty$. The equalities come
from rearranging the $\max$ and $\min$ operators and elementary implications.
By applying \Lemma{4.3} of \cite{AgrawalTeneketzisAnantharam89a}, quoted as
\Lemma{\ref{lem:quoted-lemma4-3}} below, we have {\footnotesize
\begin{eqnarray}
&&\limsup_{t\rightarrow\infty}\frac{1}{\log t+2\log\log
t}\cdot\nonumber\\
&&{\mathsf E}_{C_0}\left\{\max_{\theta>\theta_2}\sum_{s=1}^\infty
1\{\inf_{n\geq s}
    \prod_{m=1}^n\frac{F_{\theta_1}
    (dY_{m|x^*}^{1}|x^*(C_0))}{F_{\theta}
    (dY_{m|x^*}^{1}|x^*(C_0))}\leq
    t(\log t)^2\}\right\}\nonumber\\
&&\leq~\frac{1}{\min_{\theta>\theta_2}{\mathsf
E}_{C_0}\left\{\log\left(\frac{F_{\theta_1}
    (dY_{m|x^*}^{1}|x^*(C_0))}{F_{\theta}
    (dY_{m|x^*}^{1}|x^*(C_0))}\right)\right\}}\nonumber\\
&&=~\frac{1}{\min_{\theta>\theta_2}I(\theta_1,\theta|x^*(C_0))}=\frac{1}{\max_{x}\inf_{\theta>\theta_2}I(\theta_1,\theta|x)}\nonumber\\
&&=~\frac{1}{\inf_{\theta>\theta_2}\max_xI(\theta_1,\theta|x)},\nonumber
\end{eqnarray}}\cmpt
where the equalities come from the existence-of-saddle-points assumption. By
noting that $\log t\gg 2\log\log t$, this  completes the proof of
\Lemma{\ref{lem:term3e}}.
\end{proof}

By (\ref{eq:all-terms-X-depends-on-x}) and \Lemmas{\ref{lem:cond0} {\rm
through}~\ref{lem:term3e}}, it has been proved that for the $\{\phi_\tau\}$
described in \Algorithm{\ref{alg:X-depends-on-x}}, {\footnotesize
\begin{eqnarray}
\limsup_{t\rightarrow\infty}\frac{{\mathsf
E}_{C_0}\left\{T_{inf}(t)\right\}}{\log t}\leq \frac{1}{K_{C_0}}, ~~\forall
C_0\in{\mathbf \Theta}^2.\nonumber
\end{eqnarray}}\cmpt
\Lemma{4.3} of \cite{AgrawalTeneketzisAnantharam89a} is quoted as follows.
\begin{lemma}[\Lemma{4.3} of \cite{AgrawalTeneketzisAnantharam89a}]\label{lem:quoted-lemma4-3}
        Suppose $Y_1, Y_2,\cdots$ are i.i.d.\ r.v.'s taking values in a finite set~${\mathbf Y}$, with
        marginal mass function $p(y)$. Let $f^\theta:{\mathbf Y}\rightarrow{\mathbb R}$ be such that
        $0<{\mathsf E}_p\{f^\theta(Y_1)\}<\infty$, $\forall \theta\in\mathbf\Theta$, where $\mathbf\Theta$ is a finite set. Define
        $S^\theta_t=\sum_{\tau=1}^tf^\theta(Y_\tau)$, $L^\theta_A=\sum_{\tau=1}^\infty1\{\inf_{t\geq \tau}S^\theta_t\leq
        A\}$, and $L_A=\max_{\theta\in \mathbf\Theta}L_A^\theta$.
        Then {\footnotesize
        \begin{eqnarray}
                \limsup_{A\rightarrow\infty}\frac{{\mathsf E}_p\{L_A\}}{A}\leq\frac{1}{\min_{\theta\in\mathbf\Theta}{\mathsf E}_p\{f^\theta(Y_1)\}}.
        \end{eqnarray}}
\end{lemma}

Note: by incorporating Cram\'{e}r's theorem during the proof of this lemma in
\cite{AgrawalTeneketzisAnantharam89a}, it can be extended to continuous r.v.'s
$Y_1,Y_2,\cdots$, provided ${\mathsf E}_p\left\{|f^\theta(Y_1)|\right\}$ and
${\mathsf E}_p\left\{|f^\theta(Y_1)|^2\right\}$ are finite for all $\theta$.

%%%%%%%%%%%%%%%%%%%%%%%%%%%%%%%%%%%%%%%%%%%%%%%%%%%%%%%%%%%%%%%%%%%%%%%%%%%%%%%%%%%%%%%%%%
%
%   General case: the proof
%
%%%%%%%%%%%%%%%%%%%%%%%%%%%%%%%%%%%%%%%%%%%%%%%%%%%%%%%%%%%%%%%%%%%%%%%%%%%%%%%%%%%%%%%%%%
\section{Proof of \Theorems{\ref{thm: lbd-mixed-case}} and {\it \ref{thm: scheme-mixed-case}}\label{app:
pf-mixed-case}}
\begin{thmproof}{Proof of \Theorem{\ref{thm: lbd-mixed-case}} ($\log t$ lower bound):}
This proof is basically a variation of that for \Theorem{\ref{thm:
lbd-X-depends-on-x}}, with the major difference being that the competing
configuration $C'=(\theta, \theta_2)$ is now from a different set:
$\{\theta:\exists x_0, \mu_\theta(x_0)>\mu_{\theta_2}(x_0)\}$. We can first
follow line by line in the proof of \Theorem{\ref{thm: lbd-X-depends-on-x}},
and replace (\ref{eq:thm5-3}) with the following inequality.{\footnotesize
\begin{eqnarray}
&& {\mathsf E}_{C'}\left\{T_{inf}(t)\right\}\nonumber\\
&&\geq~  {\mathsf
 E}_{C'}\left\{\sum_{\tau=1}^t1\{\phi_\tau=2, M_{C'}(X_\tau)=1\}\right\}\nonumber\\
&&=~  {\mathsf
 E}_{C'}\left\{\sum_{\tau=1}^t1\{M_{C'}(X_\tau)=1\}-\right.\nonumber\\
 &&~~~~~~~~~~~~~~\left.\sum_{\tau=1}^t1\{\phi_\tau=1, M_{C'}(X_\tau)=1\}\right\}\nonumber\\
&&\geq~  {\mathsf
 E}_{C'}\left\{\sum_{\tau=1}^t1\{M_{C'}(X_\tau)=1\}-\sum_{\tau=1}^t1\{\phi_\tau=1\}\right\}\nonumber\\
%&&=~  {\mathsf
% E}_{C'}\left\{t{\mathsf P}_G(M_{C'}(X_\tau)=1)-T_1(t)\right\}\nonumber\\
&&\stackrel{(b)}{=}~  {\mathsf
 E}_{C'}\left\{\pi t-T_{1}(t)\right\}\nonumber\\
&&\stackrel{(c)}{\geq}~ \left(\pi t-\frac{\log
t}{(1+2\delta)K_{C'}}\right){\mathsf P}_{C'}\left(A_1
\right)\nonumber\\
&&\stackrel{(d)}{\geq}~\left(\pi t-\frac{\log
t}{(1+2\delta)K_{C'}}\right){\mathsf
P}_{C'}\left(A_1\cap A_2 \right) \nonumber\\
&&\stackrel{(e)}{\geq}~ \left(\pi t-\frac{\log
t}{(1+2\delta)K_{C'}}\right)e^{-\frac{(1+\delta)\log
t }{1+2\delta}}{\mathsf P}_{C_0}\left(A_1\cap A_2 \right) \nonumber\\
&&\stackrel{(f)}{=}~ {\mathcal
O}\left(t^\frac{\delta}{1+2\delta}\right),\nonumber
\end{eqnarray}}\cmpt
where the first inequality comes from dropping the other half of the events
where $\{\phi_\tau=1, M_{C'}(X_\tau)=2\}$. The second inequality comes from
dropping the condition $M_{C'}(X_\tau)=1$. With $\pi:={\mathsf
P}_G(M_{C'}(X_\tau)=1)>0$, recalling that $\theta'$ satisfies that $\exists
x_0$, such that $M_{C'}(x_0)=1$, we obtain $(b)$. $(c)$--$(f)$ follow from the
same reasoning as discussed in connection with (\ref{eq:thm5-3}). From the
contradiction of the uniformly-good-rule assumption, we have
\begin{eqnarray}
\lim_{t\rightarrow\infty} {\mathsf
P}_{C_0}\left(T_1(t)\geq\frac{(1-\epsilon)\log t}{K_{C'}}\right)=0, ~~\forall
\epsilon>0.\nonumber
\end{eqnarray}
By choosing the $\theta$ in $C'=(\theta,\theta_2)$ with the minimizing
configuration $\inf_{\{\theta:\exists x, \st
\mu_\theta(x)>\mu_{\theta_2}(x)\}}\sup_xI(\theta_1,\theta|x)$, the proof of the
first statement in \Theorem{\ref{thm: lbd-mixed-case}} follows. The second
statement in \Theorem{\ref{thm: lbd-mixed-case}} can be obtained by simply
applying Markov's inequality and the first statement.
\end{thmproof}

%%%%%%%%%%%%%%%%%%%%%%%%%%%%%%%%%%%%%%%%%%%%%%%%%%%%%%%%%%%%%%%%%%%%%%%%%%%%%%%%%%%%%%%%%%
%
%   The analysis of the scheme, general case
%
%%%%%%%%%%%%%%%%%%%%%%%%%%%%%%%%%%%%%%%%%%%%%%%%%%%%%%%%%%%%%%%%%%%%%%%%%%%%%%%%%%%%%%%%%%
\begin{thmproof}{Proof of \Theorem{\ref{thm: scheme-mixed-case}} (bound-achieving scheme):}
Following the same path as in the proof of \Theorem{\ref{thm:
scheme-X-depends-on-x}}, we first decompose the inferior sampling time instants
into disjoint subsequences, each of which will be discussed separately.
{\footnotesize
\begin{eqnarray}
&&T_{inf}(t)\nonumber\\
&&=~\sum_{\tau=1}^t1\{\phi_\tau\neq M_{C_0}(X_\tau)\}\nonumber\\
&&=~\sum_{\tau=1}^t1\{\phi_\tau\neq M_{C_0}(X_\tau), \cd{0}(\tau)\}\nonumber\\
&&~~~~+\sum_{\tau=1}^t1\{\phi_\tau\neq M_{C_0}(X_\tau), \cd{1}(\tau)\}\nonumber\\
&&~~~~+\sum_{\tau=1}^t1\{\phi_\tau\neq M_{C_0}(X_\tau), \cd{2}(\tau)\}\nonumber\\
&&~~~~+\sum_{\tau=1}^t1\{\phi_\tau\neq M_{C_0}(X_\tau), \cd{2.5}(\tau)\}\nonumber\\
&&~~~~+\sum_{\tau=1}^t1\{\phi_\tau\neq M_{C_0}(X_\tau), \hat{C}_{\tau-1}=(\theta^\alpha,\theta^\beta), \nonumber\\
&&~~~~~~~~~~~~~~~\theta^\alpha\prec\theta^\beta\neq\theta_2,\cd{3}(\tau)\}\nonumber\\
&&~~~~+\sum_{\tau=1}^t1\{\phi_\tau\neq M_{C_0}(X_\tau), \hat{C}_{\tau-1}=(\theta^\alpha,\theta^\beta), \nonumber\\
&&~~~~~~~~~~~~~~~\theta_1\neq \theta^\alpha\succ\theta^\beta,\cd{3}(\tau)\}\nonumber\\
&&~~~~+\sum_{\tau=1}^t1\{\phi_\tau\neq M_{C_0}(X_\tau), \hat{C}_{\tau-1}=(\theta^\alpha,\theta^\beta), \nonumber\\
&&~~~~~~~~~~~~~~~\theta_1\neq \theta^\alpha\prec\theta^\beta=\theta_2,\cd{3}(\tau)\}\nonumber\\
&&~~~~+\sum_{\tau=1}^t1\{\phi_\tau\neq M_{C_0}(X_\tau), \hat{C}_{\tau-1}=(\theta^\alpha,\theta^\beta), \nonumber\\
&&~~~~~~~~~~~~~~~\theta_1=\theta^\alpha\succ\theta^\beta\neq\theta_2,\cd{3}(\tau)\}\nonumber\\
&&~~~~+\sum_{\tau=1}^t1\{\phi_\tau\neq M_{C_0}(X_\tau),
\hat{C}_{\tau-1}=(\theta^\alpha,\theta^\beta)=C_0=(\theta_1,\theta_2),\nonumber\\
&&~~~~~~~~~~~~~~~\cd{3}(\tau)\}.\label{eq:all-terms-mixed-case}
\end{eqnarray}}

By exactly the same analysis as in \Lemmas{\ref{lem:cond0} {\rm
and}~\ref{lem:cond1}}, the first two sums in (\ref{eq:all-terms-mixed-case}),
concerning \cd{0} and \cd{1}, have bounded expectations. Let $\bar{\theta}$
denote the configuration satisfying $\forall x_0,
\mu_{\theta}(x_0)\leq\mu_{\bar{\theta}}(x_0)$. For the sum concerning \cd{2},
$\{\phi_\tau\neq M_{C_0}(X_\tau), \cd{2}(\tau)\}$ implies it is either
$\theta^\alpha=\bar{\theta}\neq \theta_1$ or $\theta^\beta=\bar{\theta}\neq
\theta_2$, where $(\theta^\alpha,\theta^\beta)=\hat{C}_{\tau-1}$. Both of the
above cases are discussed in \Lemma{\ref{lem:cond2}} and are proved to have
finite expectations.

For future reference, we denote the five different sums concerning \cd{3} as
{\sf term3a}, {\sf term3b}, {\sf term3c}, {\sf term3d}, and {\sf term3e}, in
order. By \Lemma{\ref{lem:term3a}} and Corollary~\ref{lem:term3b}, both {\sf
term3a} and {\sf term3b} have bounded expectations.

 If the underlying $C_0$ is not implicitly revealing, by \Lemmas{\ref{lem:term3c} {\rm and}~\ref{lem:term3d}}, {\sf
term3c} and {\sf term3d} have bounded expectation. And by
\Lemma{\ref{lem:term3e}}, $\limsup_t\frac{{\mathsf E}_{C_0}\{\mbox{\sf
term3e}\}}{\log t }\leq K_{C_0}$.

If the underlying $C_0$ is implicitly revealing, $\mbox{\sf term3e}=0$. For
{\sf term3c} and {\sf term3d}, we have {\scriptsize
\begin{eqnarray}
&&\sum_{\tau=1}^t1\{\phi_\tau\neq M_{C_0}(X_\tau), \hat{C}_{\tau-1}=(\theta^\alpha,\theta^\beta), \nonumber\\
&&~~~~~~~~~~\theta_1\neq \theta^\alpha\prec\theta^\beta=\theta_2,\cd{3}(\tau)\}\nonumber\\
&&~~~+\sum_{\tau=1}^t1\{\phi_\tau\neq M_{C_0}(X_\tau), \hat{C}_{\tau-1}=(\theta^\alpha,\theta^\beta), \nonumber\\
&&~~~~~~~~~~~~~~~\theta_1=\theta^\alpha\succ\theta^\beta\neq \theta_2,\cd{3}(\tau)\}\nonumber\\
&&\leq~\sum_{\tau=1}^t1\{\phi_\tau=1, \hat{C}_{\tau-1}=(\theta^\alpha,\theta^\beta), \theta_1\neq \theta^\alpha\prec\theta^\beta=\theta_2,\cd{3}(\tau)\}\nonumber\\
&&~~~+\sum_{\tau=1}^t1\{\phi_\tau=2, \hat{C}_{\tau-1}=(\theta^\alpha,\theta^\beta), \theta_1\neq \theta^\alpha\prec\theta^\beta=\theta_2,\cd{3}(\tau)\}\nonumber\\
&&~~~+\sum_{\tau=1}^t1\{\phi_\tau=1, \hat{C}_{\tau-1}=(\theta^\alpha,\theta^\beta), \theta_1=\theta^\alpha\succ\theta^\beta\neq \theta_2,\cd{3}(\tau)\}\nonumber\\
&&~~~+\sum_{\tau=1}^t1\{\phi_\tau=2,
\hat{C}_{\tau-1}=(\theta^\alpha,\theta^\beta),
\theta_1=\theta^\alpha\succ\theta^\beta\neq
\theta_2,\cd{3}(\tau)\},\nonumber\\
&&\label{eq:mc-tm3cd}
\end{eqnarray}
}

\noindent which is obtained by replacing the condition $\phi_\tau\neq
M_{C_0}(X_\tau)$ with either $\phi_\tau=1$ or $\phi_\tau=2$. By
\Lemma{\ref{lem:term3c}}, both the first and the fourth sums in
(\ref{eq:mc-tm3cd}) have bounded expectations. By \Lemma{\ref{lem:term3d}},
both the second and the third sums in (\ref{eq:mc-tm3cd}) also have bounded
expectations.

Note: in the proofs of \Lemmas{\ref{lem:term3c}, \ref{lem:term3d}, {\rm
and}~\ref{lem:term3e}}, there are summations or minima taken on the set
$\{\theta>\theta^\beta\}$. All those sets could be replaced by
$\{\theta:\exists x_0, \st \mu_\theta(x_0)>\mu_{\theta^\beta}(x_0)\}$ and the
rest of the proofs still follow.

We have discussed all sub-sums in (\ref{eq:all-terms-mixed-case}) except the
sum regarding \cd{2.5}. It remains to show that the sum concerning \cd{2.5} has
bounded expectation, which is addressed in the following lemma.

\begin{lemma}\label{lem:cond25}
Consider the $\{\phi_\tau\}$ described in \Algorithm{\ref{alg:mixed-case}}. For
all possible $C_0$, we have
\begin{eqnarray}
\lim_{t\rightarrow\infty}{\mathsf
E}_{C_0}\left\{\sum_{\tau=1}^t1\{\phi_\tau\neq M_{C_0}(X_\tau),
\cd{2.5}(\tau)\}\right\}<\infty.\nonumber
\end{eqnarray}
\end{lemma}

\begin{proof}
\begin{eqnarray}
&&\sum_{\tau=1}^t1\{\phi_\tau\neq M_{C_0}(X_\tau), \cd{2.5}(\tau)\}\nonumber\\
&&\leq~\sum_{(\theta,\vartheta):(\theta,\vartheta)\neq C_0}\sum_{\tau=1}^t1\{\hat{C}_{\tau-1}=(\theta,\vartheta), \cd{2.5}(\tau)\}\nonumber\\
        &&=~\sum_{(\theta,\vartheta):(\theta,\vartheta)\neq C_0}\sum_{\tau=1}^t1\{\hat{C}_{\tau-1}=(\theta,\vartheta), \cd{2.5}(\tau),\nonumber\\
        &&~~~~~~~~~~~~~~~ L_X(\tau|\cd{2.5})\in\delta\nbd(G)\}\nonumber\\
            &&~~~~+\sum_{(\theta,\vartheta):(\theta,\vartheta)\neq C_0}\sum_{\tau=1}^t 1\{\hat{C}_{\tau-1}=(\theta,\vartheta), \cd{2.5}(\tau),\nonumber\\
            &&~~~~~~~~~~~~~~~ L_X(\tau|\cd{2.5})\notin\delta\nbd(G)\}\label{eq:tm25-2}
\end{eqnarray}
By Sanov's theorem on finite alphabets, each term in the second sum is
exponentially upper bounded w.r.t.\ $\tau$, which implies that the second sum
has finite expectation. For the first sum, we have
\begin{eqnarray}
&&\sum_{\tau=1}^t1\{\hat{C}_{\tau-1}=(\theta,\vartheta), \cd{2.5}(\tau), \nonumber\\
&&~~~~~~~~~~~~~~~L_X(\tau|\cd{2.5})\in\delta\nbd(G)\}\nonumber\\
       &\leq&\sum_{\tau=1}^t1\{\hat{C}_{\tau-1}=(\theta,\vartheta), \theta\neq \theta_1,\cd{2.5}(\tau),\nonumber\\
       &&~~~~~~~~~~~~~~~ L_X(\tau|\cd{2.5})\in\delta\nbd(G)\}\nonumber\\
            &&+\sum_{\tau=1}^t1\{\hat{C}_{\tau-1}=(\theta,\vartheta), \vartheta\neq\theta_2, \cd{2.5}(\tau), \nonumber\\
            &&~~~~~~~~~~~~~~~L_X(\tau|\cd{2.5})\in\delta\nbd(G)\},\label{eq:tm25-3}
\end{eqnarray}
which is obtained by considering whether $\theta\neq\theta_1$ or
$\vartheta\neq\theta_2$, recalling that $(\theta,\vartheta)\neq C_0$. Since
these two sums are symmetric, henceforth we show only the finite expectation of
the first sum in (\ref{eq:tm25-3}). The finite expectation of the second sum
then follows by symmetry. {\footnotesize
\begin{eqnarray}
&&       \sum_{\tau=1}^t1\{\hat{C}_{\tau-1}=(\theta,\vartheta), \theta\neq \theta_1,\cd{2.5}(\tau), \nonumber\\
&&~~~~~~~~~~~~~~~L_X(\tau|\cd{2.5})\in\delta\nbd(G)\}\nonumber\\
        &&\leq~\sum_{\tau=1}^t1\{\exists x,\st M_{(\theta,\vartheta)}(x)=1,
            \rho(L_1^x(\tau-1),F_{\theta_1}(\cdot|x))>\epsilon, \nonumber\\
            &&~~~~~~~~~~~~\cd{2.5}(\tau),L_X(\tau|\cd{2.5})\in\delta\nbd(G)\}\nonumber\\
        &&\leq~                \sum_{x:M_{(\theta,\vartheta)}(x)=1}
                \sum_{\tau'=1}^\infty1\{\exists n\geq [\tau'{\mathsf P}_G(X=x)(1-\delta)],
                \st\nonumber\\
&&~~~~~~~~~~~~~~~
\rho(L_1^x(n),F_{\theta_1}(\cdot|x))>\epsilon\}\label{eq:tm25-1}
\end{eqnarray}}

The first inequality comes from the definition of \cd{2.5}: since
$\hat{C}_{\tau-1}=(\theta,\vartheta)$ is implicitly revealing, there must be an
$x$ \st $M_{\hat{C}_{\tau-1}}=1$. And since the estimate $\theta\neq\theta_1$,
for that specific $x$, the distance between $L_1^x$ and $F_{\theta_1}(\cdot|x)$
must be greater than $\epsilon$. The second inequality comes from changing the
time index to $\tau'$, the time instants  at which $X_s=x$ and \cd{2.5} is
satisfied, and extending the summation to infinity. (This change of the time
index is similar to the one described in
(\ref{eq:change-time1})--(\ref{eq:change-time2})).

Thus by Sanov's theorem on $\mathbb R$, the expectation of each term in
(\ref{eq:tm25-1}) is exponentially upper bounded w.r.t.\ $\tau'$, which implies
finite expectation of the entire sum in (\ref{eq:tm25-1}). By the discussions
on (\ref{eq:tm25-2}), (\ref{eq:tm25-3}), and (\ref{eq:tm25-1}),
\Lemma{\ref{lem:cond25}} is proved.
\end{proof}

 From the above discussion of the sub-sums in (\ref{eq:all-terms-mixed-case}), we conclude that the modified
 scheme, $\{\phi_\tau\}$ in \Algorithm{\ref{alg:mixed-case}}, has bounded ${\mathsf
 E}_{C_0}\{T_{inf}(t)\}$ if the underlying $C_0$ is implicitly revealing.
If $C_0$ is not implicitly revealing, the $\{\phi_\tau\}$ in
\Algorithm{\ref{alg:mixed-case}} achieves the new $\log t$ lower bound
(\ref{eq: lbd-mixed-case}).
\end{thmproof}

% use section* for acknowledgement
%\section*{Acknowledgment}
% optional entry into table of contents (if used)
%\addcontentsline{toc}{section}{Acknowledgment}
%The authors would like to thank...

% trigger a \newpage just before the given reference
% number - used to balance the columns on the last page
% adjust value as needed - may need to be readjusted if
% the document is modified later
%\IEEEtriggeratref{8}
% The "triggered" command can be changed if desired:
%\IEEEtriggercmd{\enlargethispage{-5in}}

% references section
% NOTE: BibTeX documentation can be easily obtained at:
% http://www.ctan.org/tex-archive/biblio/bibtex/contrib/doc/

% can use a bibliography generated by BibTeX as a .bbl file
% standard IEEE bibliography style from:
% http://www.ctan.org/tex-archive/macros/latex/contrib/supported/IEEEtran/bibtex
%\bibliographystyle{IEEEtran.bst}
% argument is your BibTeX string definitions and bibliography database(s)
%\bibliography{IEEEabrv,../bib/paper}
%
% <OR> manually copy in the resultant .bbl file
% set second argument of \begin to the number of references
% (used to reserve space for the reference number labels box)

\bibliography{bandit}
\bibliographystyle{IEEEtran}

% biography section
%
% If you have an EPS/PDF photo (graphicx package needed) extra braces are
% needed around the contents of the optional argument to biography to prevent
% the LaTeX parser from getting confused when it sees the complicated
% \includegraphics command within an optional argument. (You could create
% your own custom macro containing the \includegraphics command to make things
% simpler here.)
%\begin{biography}[{\includegraphics[width=1in,height=1.25in,clip,keepaspectratio]{mshell}}]{Michael Shell}
% where an .eps filename suffix will be assumed under latex, and a .pdf suffix
% will be assumed for pdflatex; or if you just want to reserve a space for
% a photo:

\begin{biography}{Chih-Chun Wang} received the B.E. degree in electrical engineering from National Taiwan University, Taipei, Taiwan in 1999.
He is currently working toward the Ph.D. degree in electrical engineering at
Princeton University, Princeton, NJ. He worked in COMTREND Corporation, Taipei,
Taiwan, from 1999-2000, and spent the summer of 2004 with Flarion Technologies.
His research interests are in optimal control, information theory and coding
theory, especially on iterative decoding of LDPC codes.
\end{biography}
\begin{biography}{Sanjeev R. Kulkarni}
(M'91, SM'96, F'04) received the B.S. in Mathematics, B.S.\ in E.E., M.S.\ in
Mathematics from Clarkson University in 1983, 1984, and 1985, respectively, the
M.S.\ degree in E.E.\ from Stanford University in 1985, and the Ph.D. in E.E.
from M.I.T.\ in 1991. \\From 1985 to 1991 he was a Member of the Technical
Staff at M.I.T.\ Lincoln Laboratory working on the modeling and processing of
laser radar measurements. In the spring of 1986, he was a part-time faculty at
the University of Massachusetts, Boston.  Since 1991, he has been with
Princeton University where he is currently Associate Professor of Electrical
Engineering and Associate Dean of Academic Affairs in the School of Engineering
and Applied Science.  He spent January 1996 as a research fellow at the
Australian National University, 1998 with Susquehanna International Group, and
summer 2001 with Flarion Technologies.

Prof. Kulkarni received an ARO Young Investigator Award in 1992, an NSF Young
Investigator Award in 1994, and several teaching awards at Princeton
University.  He has served as an Associate Editor for the IEEE Transactions on
Information Theory.  Prof.\ Kulkarni's research interests include statistical
pattern recognition, nonparametric estimation, learning and adaptive systems,
information theory, wireless networks, and image/video processing.
\end{biography}

% insert where needed to balance the two columns on the last page
%\newpage

\begin{biography}{H. Vincent Poor}
(S'72, M'77, SM'82, F'87) received the Ph.D. degree in EECS in 1977 from
Princeton University, where he is currently the George Van Ness Lothrop
Professor in Engineering.  From 1977 until he joined the Princeton faculty in
1990, he was a faculty member at the University of Illinois at
Urbana-Champaign. His research interests are primarily in the areas of
stochastic analysis and statistical signal processing, with applications in
wireless communications and related areas.  Among his publications in this area
is the recent book, {\it Wireless Networks: Multiuser Detection in Cross-Layer
Design}.

Dr. Poor is a member of the National Academy of Engineering, and is a Fellow of
the Institute of Mathematical Statistics, the Optical Society of America, and
other organizations. In 1990, he served as the President of the IEEE
Information Theory Society and he is currently the Editor-in-Chief of the {\it
IEEE Transactions on Information Theory}.  Among his recent honors are the
Joint Paper Award of the IEEE Communications and Information Theory Societies
(2001), the NSF Director's Award for Distinguished Teaching Scholars (2002), a
Guggenheim Fellowship (2002-03), the IEEE EAB Major Educational Innovation
Award (2004), and the IEEE Education Medal (2005).

\end{biography}

% You can push biographies down or up by placing
% a \vfill before or after them. The appropriate
% use of \vfill depends on what kind of text is
% on the last page and whether or not the columns
% are being equalized.

%\vfill

% Can be used to pull up biographies so that the bottom of the last one
% is flush with the other column.
%\enlargethispage{-5in}

% that's all folks
\end{document}